\documentclass[%
preprint,onecolumn,
amsmath,amssymb,
aps,
]{revtex4-1}
\usepackage{natbib}
\usepackage{subfigure}
\usepackage{amsthm}
\usepackage[normalem]{ulem}
\usepackage{graphicx}
\usepackage{dcolumn}
\usepackage{color}
\usepackage{multirow}
\usepackage[table]{xcolor}
\usepackage{colortbl}
\usepackage[colorlinks=true,linkcolor=blue,
urlcolor=blue,citecolor=blue]{hyperref}

\begin{document}

\title{Resonant triad interactions in stably-stratified uniform shear flow}

\author{Lima Biswas\footnote{ma16d003@smail.iitm.ac.in} and Priyanka Shukla\footnote{priyanka@iitm.ac.in}}
\affiliation{Department of Mathematics,\\
Indian Institute of Technology Madras,\\
Chennai 600036, India
}

\begin{abstract}
We investigate exact and near resonant triad interactions (RTI) in a two-dimensional stably stratified uniform shear flow confined between two infinite parallel walls in the absence of viscous and diffusive effects. RTI occur when three interacting waves satisfy the resonance conditions of the form $k_1\pm k_2=k_3$ and $\omega_1\pm \omega_2=\omega_3$ with $k_i$ and $\omega_i$ being the wavenumber and frequency of the $i^\mathrm{th}$ wave ($i \in \{1, 2, 3\})$, respectively. The linear stability problem is solved analytically, which gives the eigenfunctions in the form of the modified Bessel functions. It is identified that an interaction between two primary modes having the same frequency $\omega$ but different wavenumbers $k_m$ and $k_n$ produces two different secondary modes: one time-dependent (superharmonic) mode having frequency $2\omega$ and wavenumber $k_m+k_n$, and the other time-independent (subharmonic) mode with $\omega=0$ and wavenumber $k_m-k_n$. The differential equation governing the spatial amplitude of the superharmonic mode is solved numerically as well as analytically using the method of variation of parameters. It turns out that the linear operator associated with the differential equation of the superharmonic mode is the same as the linear stability operator and that the solvability condition of the differential equation is found to be associated with the existence of RTI. The existence of resonant triad interactions predicted by the dispersion relation, are justified by showing the divergence of the spatial amplitude of superharmonic mode. Various cases of wave interactions in a stably stratified shear flow are analysed in the presence of a resonant triad for various frequencies and linear stratifications. 
\end{abstract}

\maketitle

\section{Introduction}

Stratified flows, i.e.~flows in which density depends upon the gravity, are ubiquitous in nature, for instance, in lakes, rivers, Earth's ocean and its atmosphere, etc. With respect to gravity, stratification can be classified into two categories, namely, stable and unstable stratifications. When the density increases (or decreases) in the direction of gravity, the flow is defined as the stably (or unstably) stratified flows. Stable and unstable stratified flows in bounded and unbounded geometries have been studied from decades~\cite{Yih1969,bell_1975,Grimshaw2002,PW2003}. Stably stratified flows are of great interest from a geophysical point of view since various kind of waves exist at the surface and at the internal parts of these flows, for instance, internal waves, oceanic surface waves, solitons, gravity waves, Rossby waves, acoustic-gravity waves, tsunami waves, etc.~\citep{Lighthill2001,loper_2017}. Notably, there are a number of instability-induced phenomena associated with these waves in stably stratified flows; which are of interest from engineering as well as geophysical point of views, however, not all of them are fully understood at present~\citep{Turner1973,SS2002}. One of the well-known phenomena, associated with the internal waves in stably stratified flows, is resonant triad interactions (RTI); which arise due to a nonlinear wave interaction satisfying the resonance conditions given by $k_1+k_2=k_3$ and $\omega_1+\omega_2=\omega_3$, where $k_1$, $k_2$ and $k_3$ are the wavenumbers and $\omega_1$, $\omega_2$ and $\omega_3$ are the frequencies of the interacting waves. It is worthwhile to note that  RTI---being the underlying energy transfer strategy in internal waves generated, for instance, by winds and tides in the ocean---play an important role in ocean mixing, ocean internal tides, etc. If  small wavenumber or frequency mismatch occurs in the condition of exact RTI, i.e. $k_1+k_2+k_3 = \delta k$ and $\omega_1+\omega_2+\omega_3=\delta \omega$, the interaction is defined as the near resonance triad interaction, which was first developed by~\citet{Armstrong_1962} for the interaction among light waves. In geophysical context, such interactions are studied in by~\citet{benney_1962,koudella_staquet_2006,craik_book,lamb_2007}.

There are two significant consequences of RTI, namely, the triadic resonance instability (TRI) and the superharmonic wave generation~\citep{McComasBretherton1977,DiasChristian1999,alam_liu_yue_2011}. In the former, one parent (primary) wave with wavenumber $k_0$ and frequency $\omega_0$ generates two sibling (secondary) waves with wavenumbers $k_\pm$ and frequencies $\omega_\pm$. These two sibling waves form a resonant triad with the parent wave, and due to this, there exists a continuous transfer of energy from the parent wave to the sibling waves~\citep{PRF_TRI_sutherland,Ann_Rev_TRI}. A special type of TRI is the parametric subharmonic instability, where 
the wavenumber and frequency of each sibling wave is exactly half of that of  the parent wave, i.e $k_\pm = k_0/2$ and $\omega_\pm=\omega_0/2$~\citep{koudella_staquet_2006,bourget_dauxois_joubaud_odier_2013}. In the latter, two primary internal waves having fixed equal frequencies generate an internal wave with the frequency equal to twice of the frequency of a primary wave~\citep{wunsch_2017,varma_mathur_2017}.  

If should be noted that the internal waves appear spontaneously in stably stratified flows even for very small  disturbances. Thanks to the pioneering works of~\citet{Taylor1931} and~\citet{Goldstein1931}, the linear stability of stratified flows is no longer unclear. \citet{Taylor1931} and~\citet{Goldstein1931} conjectured that the sufficient condition for the stability of a heterogeneous shear flow is ${\rm Ri}>0.25$, where ${\rm Ri}$ is the Richardson number (defined as the ratio of the squared buoyancy frequency to the square of vertical shear), and that all stable modes are neutral. The work of~\citet{Taylor1931} has been extended by~\citet{eliassen1953two} wherein the authors analysed a stratified shear flow with a constant density gradient in a bounded geometry by formulating an initial value problem. Their analysis revealed two important results. Firstly, the perturbation behaves asymptotically like $t^{-0.5+(0.25-{\rm Ri})^{1/2}}$ for all $-0.75<{\rm Ri}<0.25$, where $t$ denotes the time, and hence the flow remains unstable for $-0.75<{\rm Ri}<0$. Secondly, the flow would be exponentially unstable for ${\rm Ri}<-0.75$. Following Taylor's and Goldstein's works,~\citet{Case1960} investigated the stability of a stably stratified shear flow for ${\rm Ri}>0$ in an idealized atmosphere by considering the inertial effects of the density in the flow. He showed that (i) there exists an infinite number of discrete stable eigenvalues for ${\rm Ri}>0.25$, (ii) perturbation vanishes asymptotically as $t^{-1/2}$, and (iii) all modes are neutral. A year later,~\citet{miles_1961} proved  Taylor's conjecture mathematically and also analysed the results of~\cite{eliassen1953two} and~\cite{Case1960}. 

Over the past many decades a surge of research involved to answer the question of nonlinear wave interactions with or without resonance, however, the credit for developing the theory of nonlinear wave interactions and resonance goes to the seminal work by~\citet{phillips_1960} who studied the interaction of finite-amplitude surface gravity waves in deep water. Using the perturbation method, he solved the nonlinear water wave equation and observed a continuous transfer of energy under the resonance condition. He also showed that no second-order resonant interactions are possible among a triad of surface gravity waves, nevertheless, the cubic order resonant interactions can occur among a group of four surface gravity waves. The prediction of~\citet{phillips_1960} investigation has also been verified experimentally by~\citet{longuet-higgins_1962}. Extending Phillips's theory, \citet{benney_1962} demonstrated the energy-sharing mechanism among surface-gravity waves under the exact and near resonance conditions, and found that there is a direct but slow exchange of energy among the order-one modes. In contrast to a homogeneous single-layer fluid~\cite{phillips_1960},~\citet{Ball1964} showed that the second-order resonant interaction leading to the RTI is indeed possible between surface and interfacial-gravity waves in two-layer homogeneous fluid systems. For a comprehensive review on  RTI among surface waves, the reader is referred to~\cite{Hammack1993}.


Until~\citet{thorpe_1966}, nonlinear wave interactions were studied for the homogeneous (i.e. constant density) medium.~\citet{thorpe_1966} studied the interaction of waves in continuously stably stratified inviscid fluid in the absence of background flow. He showed that RTI are possible when two free-surface waves interact with an internal wave or when all three waves are internal gravity waves provided not all belong to the same mode. Later, the effect of background (or mean) flow on the gravity wave interactions in the inviscid fluid is addressed by~\citet{kelly_1968}. In particular,~\citet{kelly_1968} has examined the second-order resonant interaction among gravity waves in the presence of two special types of mean flow, namely the inviscid homogeneous jet and a stably stratified antisymmetric shear layer. It has then been found that the presence of mean flow (shear, pressure, etc.) not only affects the rate of energy transfer among waves but also energy transfer from the mean flow to each disturbances due to change in the Reynolds stress. In a similar study~\citet{craik_1968} proved that the uniform shear flow of the liquid having small viscosity allows strong second-order resonant interaction among three gravity waves. This resonant triad interaction results into continuous wave growth due to the transfer of energy from the mean flow. In this process the total energy of the wave increases with time. Furthermore, he studied the interaction of waves on the interface of two fluid layers of different densities and velocities. In the oceanic scenario, the vorticity wave (generated due to the mean shear flow) and the internal gravity waves participate in near or exact resonant interactions. The presence of shear current affects the nonlinear dynamics of the internal wave and the energy exchange among different harmonics of the internal waves significantly~ \citep{voronovich_pelinovsky_shrira_1998,chen_zou_2019}.~\citet{grimshaw_1988} focused on the situations where the resonant conditions can meet locally under some circumstances for the interaction of internal gravity waves propagating in a stratified shear flow, where the background shear flow vary slowly with respect to the waves. Later he extended this work to study the higher-order resonant interactions near a critical level. He proved the occurrence of explosive resonant interaction in a continuously stratified shear flow~\citep{grimshaw_1994}.

Although the second-order wave interactions in a stably stratified uniform shear flow have been studied theoretically in~\citep{thorpe_1966,craik_1968}, identification of the exact and near resonant triads---emerging from the modal interactions of internal gravity waves---as well as spatial locations in the parameter space are poorly understood.

In this context, the objective of the present work is to set up the second-order weakly nonlinear solution containing an arbitrary sum of vertical modes at a fixed frequency $\omega$ to identify the existence of  RTI.  The paper is organized as follow. The problem definition and non-dimensional equations are given in~\S\ref{sec:Problem definition}. Linear stability problem and discussions on the analytical solution are studied in~\S\ref{sec:Linear stability theory}. The nonlinear problem is discussed in \S\ref{sec:Nonlinear problem}. The first- and second-order solutions are given in~\S\S\ref{subsec:First order solution} and~\ref{subsec:2ndsol}, respectively. The analytical solution for the second-order problem is formulated in \S\ref{subsec:analytical_2ndsol}. Results are given in \S\ref{sec:Results}. At the end, conclusions are given in \S\ref{sec:Conclusion}.

\section{Problem definition}
\label{sec:Problem definition}

Consider a two-dimensional stably stratified incompressible 
flow bounded between two oppositely moving walls at $z=\pm L$ with speed $\bar{U}_w$ along the $x$-direction.
The background flow under consideration is a parallel shear flow, $\bar{U}(z) = \bar{U}_w\, z/L$, in a stably stratified medium with density satisfying 
\begin{equation}
\bar{\varrho}(z)=  \rho_m + \bar{\rho}(z)=\rho_m\left(1-\frac{z}{L}\right),
\label{eqn:density_profile}
\end{equation}
where $\rho_m$ is a constant reference density. Thus, the shear flow is superimposed in a background state such that the pressure $\bar{p}(z)$ and density $\bar{\varrho}(z)$ are in hydrostatic balance:
\begin{equation}
\frac{\partial \bar{p}}{\partial z} = - g \bar{\varrho}.
\label{eqn:hydrostatic_pressure}
\end{equation} 
Note that gradient of background density is a constant.

In general, the mass and momentum balance equations for an incompressible inviscid flow read
\begin{align}
\nabla \cdot \boldsymbol{u} &=0\quad\mbox{and} \quad
\rho \frac{{\rm D} \boldsymbol{u}}{{\rm D}t} = - \nabla  p + \rho \boldsymbol{g},
\label{eqn:balanceEq}
\end{align} 
where ${\rm D}/{\rm D}t \equiv \partial/\partial t + \boldsymbol{u}\cdot \nabla$ is the material derivative, $\boldsymbol{u} = (u,w)$ is the velocity vector, $\rho$, $p$ and $\boldsymbol{g}$ are the density, pressure and acceleration due to gravity, respectively. In addition, the incompressibility of fluid particle leads to a density equation
\begin{equation}
\frac{{\rm D}\rho}{{\rm D}t}=0. 
\label{eqn:density}
\end{equation}

\subsection{Nonlinear disturbance equations}

In order to find the disturbance equations each of the flow variables is decomposed into its background state and infinitesimally small disturbance, 
\begin{align}
\left.
\begin{aligned}
\rho(x,z,t) &= \rho_m+ \bar{\rho}(z)+\rho^\prime(x,z,t),\\
p(x,z,t)  &= \bar{p}(z)+p^\prime(x,y,t),\\
[u(x,z,t) ,w(x,z,t)]  &= [\bar{U}(z) + u', w'].
\end{aligned}
\label{eqn:decomposition}
\right\}
\end{align}
To simplify the problem, we assume here the Boussinesq approximation, i.e.~all density variations are small compared to $\rho_m$ so that $\bar{\rho} + \rho' << \rho_m$ except the gravity term where $\rho'/\rho_m$ is significant. Substituting above decompositions~\eqref{eqn:decomposition} into~\eqref{eqn:balanceEq}--\eqref{eqn:density} and omitting the superscript prime over the perturbed variables, we get
\begin{subequations}
\begin{gather}
\label{eqn:rho_pert}
\left(\frac{\partial}{\partial t}+\bar{U}(z)\frac{\partial}{\partial x}\right)\rho= -w\frac{\partial \bar{\rho}}{\partial z}-\left(u\frac{\partial\rho}{\partial x}+w\frac{\partial\rho}{\partial z}\right),\\
\label{eqn:continuity}
\frac{\partial u}{\partial x}+\frac{\partial w}{\partial z}=0,
\\
\left(\frac{\partial }{\partial t}+\bar{U}(z)\frac{\partial }{\partial x} \right)u+ w \frac{\partial \bar{U}(z)}{\partial z} = -\frac{1}{\rho_m}\frac{\partial p}{\partial x} - \left(u\frac{\partial u}{\partial x}+w\frac{\partial u}{\partial z}\right),
\label{eqn:pert_u}\\
\left(\frac{\partial }{\partial t}+ 
\bar{U}(z)\frac{\partial }{\partial x}\right)w = -\frac{1}{\rho_m}\frac{\partial p}{\partial z}-  
\frac{g  \rho}{\rho_m} - \left(u\frac{\partial w}{\partial x}+w\frac{\partial w}{\partial z}\right).
\label{eqn:pert_w}
\end{gather}
\end{subequations}
At the boundaries $z=\pm L$, the normal component of the velocity perturbation and the density perturbation are assumed to be zero. Eliminating the pressure by differentiating $x$~\eqref{eqn:pert_u} and $z$~\eqref{eqn:pert_w} momentum equations with respect to $z$ and $x$, respectively, and then subtracting the resulting equations; using the continuity equation~\eqref{eqn:continuity}, we get an equation for the perturbed vorticity $\eta=\partial u/\partial z-\partial w/\partial x$ as
\begin{align}
\left(
\frac{\partial }{\partial t}+\bar{U}(z)\frac{\partial }{\partial x} \right)\eta &= 
\frac{g}{{\rho}_m} \frac{\partial \rho}{\partial x} -\left(u\frac{\partial \eta}{\partial x}+w\frac{\partial \eta}{\partial z}\right).
\label{vorticity equation}
\end{align}
Note that equations~\eqref{eqn:continuity}--\eqref{eqn:pert_w} reduce to a single equation of vorticity~\eqref{vorticity equation} and therefore four disturbance equations~\eqref{eqn:rho_pert}--\eqref{eqn:pert_w} reduce to two equations for $\rho$ and $\eta$, see~\eqref{eqn:rho_pert} and~\eqref{vorticity equation}.

\subsection{Non-dimensional equations}

For non-dimensionalisation, we use half of the gap between the walls, $L$, as a reference length scale, $\rho_m$ as a reference density, the background flow speed at the wall, $\bar{U}_w$, as a reference velocity and $L/\bar{U}_w$ as a reference time scale. Here onward all the variables are in dimensionless form. As the flow is two-dimensional, one can introduce the stream function $\psi$ such that $(u,w)=(\partial \psi/\partial z, -\partial \psi/\partial x)$ and thus $\eta=\nabla^2\psi$. In terms of the  stream function, the dimensionless disturbance equations read 
\begin{align}
&\left(\frac{\partial}{\partial t} + \bar{U}(z) \frac{\partial}{\partial x}\right)\rho =- \frac{N^2}{N_0^2}
\frac{\partial \psi}{\partial x} 
+J(\psi,\rho),
\label{eqn:dimlessrho}
\\
&\left(\frac{\partial}{\partial t} + \bar{U}(z) \frac{\partial}{\partial x}\right)\nabla^2\psi = {\rm Ri}_0 \frac{\partial \rho}{\partial x} 
+J\left(\psi,\nabla^2 \psi \right),
\label{eqn:dimlesspsi}
\end{align}
where 
$J(f,g)=\frac{\partial f}{\partial x} \frac{\partial g}{\partial z} - \frac{\partial f}{\partial z}\frac{\partial g}{\partial x}$ is the Jacobian determinant. Here, $\bar{U}(z)=z$ and $N = \sqrt{-\frac{g}{{\rho_m}} \frac{{\rm d}\bar{\rho}}{{\rm d}z}}$ is the buoyancy frequency, which is constant due to linear stratification profile~\eqref{eqn:density_profile}. The buoyancy frequency measures the atmospheric stratification, which describes the stability of a stratified fluid in terms of density variations experienced by a fluid parcel displaced along the gradient direction. While $N^2>0$ implies the stable stratification, $N^2<0$ denotes the unstable one.
The buoyancy frequency $N$ is scaled using $N_0=\sqrt{g/L}$. The ratio $N/N_0$ represents the angular frequency of the vertically displaced fluid and its square is proportional to the gradient of background density.

The other two dimensionless parameters are ${\rm Ri}_0=g\,L/\bar{U}_w^2=N_0^2\,L^2/\bar{U}_w^2$ and ${\rm Ri}(z)=\left(N\,L/\bar{U}_w\right)^2={\rm Ri}_0\,N^2/N_0^2 $, referred to as the bulk Richardson number and the local Richardson number, respectively. Physically, the Richardson number is a ratio of the squared buoyancy frequency to the square of the vertical shear. Here, the vertical gradient of the horizontal velocity has the same dimension as the dimension of the frequency. Local variation of the background density gradient is represented by the local Richardson number, which is, in general, a function of $z$, whereas the bulk Richardson number ($\rm Ri_0$) is the reference quantity. In the present paper, we focus on the stably stratified linear density variation which implies that ${\rm d}\bar{\rho}/{\rm d}z={\rm const.}<0$. Therefore $N^2/N_0^2$ always represents a positive real constant, and hence the local Richardson number $\rm Ri$ remains constant. We shall consider different uniform stable stratifications by varying the squared of buoyancy frequency ($N^2/N_0^2$) and the local Richardson number ($\rm Ri$).

Equations~\eqref{eqn:dimlessrho} and~\eqref{eqn:dimlesspsi} can also be expressed in the matrix form
\begin{equation}
\begin{bmatrix}
\left( \frac{\partial}{\partial t}+\bar{U}(z)\frac{\partial }{\partial x}\right) \nabla^2 
& -{\rm Ri}_0 \,\frac{\partial}{\partial x}
\\[.5em]
 \frac{N^2}{N_0^2}\,\frac{\partial}{\partial x} & \left( \frac{\partial}{\partial t}+\bar{U}(z)\frac{\partial }{\partial x}\right)
\end{bmatrix}
\begin{bmatrix}
\psi\\
\rho
\end{bmatrix}
=
\begin{bmatrix}
J\left(\psi,\nabla^2 \psi\right) \\
J\left(\psi,\rho \right)
\end{bmatrix}.
\end{equation}
Furthermore eliminating the density gradient term of~\eqref{eqn:dimlesspsi} using~\eqref{eqn:dimlessrho}, one obtains
\begin{equation}
\left[
\left( \frac{\partial}{\partial t}+\bar{U}(z)\frac{\partial }{\partial x}\right)^2 \nabla^2  + {\rm Ri}\, \frac{\partial^2 }{\partial x^2} 
\right]\psi
=
\left( \frac{\partial}{\partial t}+\bar{U}(z)\frac{\partial }{\partial x}\right) J\left(\psi,\nabla^2\psi\right)
+ 
{\rm Ri}_0 \frac{\partial J\left(\psi,\rho\right)}{\partial x}.
\label{eqn:tg_nonlinear}
\end{equation}
The dimensionless boundary conditions are expressed as $\psi = \rho=0$ at $z=\pm 1$.

\section{Linear stability theory}
\label{sec:Linear stability theory}

In the linear stability analysis, we neglect the nonlinear terms of disturbances and solve the resulting linear system by assuming the normal mode solution as
\begin{align}
\left(\psi(x,z,t),\rho(x,z,t)\right)=\left(\hat{\psi}(z),\hat{\rho}(z)\right)e^{ik(x-c t)},
\label{eq:normal_mode}
\end{align}
where the hat over the quantity represents the complex amplitude, $k$ is the streamwise wavenumber and $c$ is the complex phase velocity whose imaginary part ($c_i$) determines the linear stability of the flow. The base flow is stable, neutrally stable or unstable if $c_i<0$, $c_i=0$ or $c_i>0$, respectively. Neglecting the right-hand side nonlinear terms of~\eqref{eqn:dimlessrho}--\eqref{eqn:dimlesspsi} and substituting the normal mode solution~\eqref{eq:normal_mode} into~\eqref{eqn:dimlessrho}--\eqref{eqn:dimlesspsi}, we arrive at a differential system for variables $\hat{\psi}$ and $\hat{\rho}$:
\begin{subequations}
\begin{align}
\left(\bar{U}(z) -c\right) \hat{\rho} &= -\frac{N^2}{N_0^2}\,\hat{\psi},
\label{eqn:gen_evp_implicitform_a}
\\
\left(\bar{U}(z) -c\right)\left(D^2 - k^2\right) \hat{\psi} &= {\rm Ri}_0\, \hat{\rho}, 
\label{eqn:gen_evp_implicitform_b}
\end{align}
\label{eqn:gen_evp_implicitform}
\end{subequations}
where $D={\rm d}/{\rm d}z$ and $D^2={\rm d}^2/{\rm d}z^2$. Eliminating density from~\eqref{eqn:gen_evp_implicitform}, we get
\begin{equation}
\left[
\left(\bar{U}(z)-c\right)^2\left(D^2-k^2\right) + {\rm Ri}\, \right]\hat{\psi} =0,
\label{eq:tg_linear}
\end{equation}
which is the modified Taylor--Goldstein equation for the linear background shear flow with uniform stratification. The system of differential equations~\eqref{eqn:gen_evp_implicitform} or equivalently~\eqref{eq:tg_linear}, along with the boundary conditions $\hat{\psi}=\hat{\rho}=0$ at $z=\pm 1$ forms a generalized eigenvalue problem.

It is worth noticing that the linear stability problem~\eqref{eqn:gen_evp_implicitform} or~\eqref{eq:tg_linear} becomes singular when the background shear flow $\bar{U}(z)=z \in [-1,1]$ coincides with the phase velocity $c$ of the perturbation, which leads to all continuous modes in the spectrum. It is verified that for $\rm Ri > 0.25$, the phase speed $c$ does not match with $\bar{U}(z)=z$ in the flow domain, hence all eigenvalues of \eqref{eqn:gen_evp_implicitform} corresponding to each wavenumber are discrete. The choice of $\rm Ri > 0.25$ also ascertains that the system~\eqref{eqn:gen_evp_implicitform} is neutrally stable, i.e. all eigenvalues for every wavenumber are real~\cite{Case1960}. This means that $c$ takes values either greater than $1$ or less than $-1$. At fixed $k$, if $\hat{\psi}(z)$ is an eigenfunction corresponding to the eigenvalue c then $\hat{\psi}(-z)$ is an eigenfunction  corresponding to the eigenvalue $-c$~\citep{davey_reid_1977}. Thus it is sufficient to consider $c>1$ in the present problem. Interestingly, the eigenvalue problem~\eqref{eq:tg_linear} admits an analytical solution in terms of the modified Bessel function~\citep{Engevik1971,Engevik1973,eliassen1953two}, see~\S\ref{subsec:analytical_sol}.

\subsection{Analytical solution and dispersion relation}
\label{subsec:analytical_sol}

The second-order differential equation~\eqref{eq:tg_linear} can be transformed into the modified Bessel equation 
\begin{equation}
p^2\,\frac{{\rm d}^2 \phi}{{\rm d} p^2}+p\, \frac{{\rm d}\phi}{{\rm d}p}-\left(p^2+\nu^2\right)\,\phi = 0,
\label{eqn:psi_Bessels_form}
\end{equation}
where
$\hat{\psi}(z)= \sqrt{p(z)}\,\phi(p(z))$, $p(z)=k\,(z-c)$ and $\nu= \sqrt{0.25-{\rm Ri}}$, 
with ${\rm Ri} >0.25$ everywhere in the flow domain and therefore $\nu$ is always an imaginary number.
The general solution of the modified Bessel equation~\eqref{eqn:psi_Bessels_form} has the form
\begin{equation}
\phi(p) = AI_\nu(p) + B K_\nu(p), 
\end{equation}
where $A$ and $B$ are two arbitrary constants and $I_\nu(p)$ and $K_\nu(p)$ are the modified Bessel functions of order $\nu$ in variable $p$. Consequently, the general solution of \eqref{eq:tg_linear} reads
\begin{equation}
\hat{\psi}(z)=A f_1(z) + B f_2(z),
\label{eqn:gen_sol_psi}
\end{equation}
where 
\begin{align}
f_1(z) &= \sqrt{k(z-c)} I_\nu(k(z-c)) \quad\mbox{and}\quad 
f_2(z) = \sqrt{k(z-c)} K_{\nu}(k(z-c)),
\end{align}
are the two linearly independent solutions of the Taylor-- Goldstein equation~\eqref{eq:tg_linear}.

For finding the arbitrary constants $A$ and $B$, we apply the boundary conditions $\hat{\psi}(\pm1)=0$  in the general solution~\eqref{eqn:gen_sol_psi}, which yields
\begin{subequations}
\label{eqn:AB}
\begin{align}
A I_{\nu}\left(k(-1-c)\right)+ B K_{\nu}\left(k(-1-c)\right)&=0,
\label{eqn:AB_a}
\\
A I_{\nu}\left(k(1-c)\right)+ B K_{\nu}\left(k(1-c)\right)&=0. 
\label{eqn:AB_b}
\end{align}
\end{subequations}
The condition for the existence of non-trivial solutions of~\eqref{eqn:AB} yields the dispersion relation 
\begin{equation}
\mathcal{D}\left(c,\,k; \,\rm Ri\right)=\left|
\begin{array}{ll}
I_\nu(k(-1-c)) & K_{\nu}(k(-1-c)) \\
I_\nu(k(1-c))  & K_{\nu}(k(1-c))
\end{array}
\right|=0,
\label{eqn:desRel_kc}
\end{equation}
or in terms of frequency $\omega=c \, k$
\begin{equation}
\mathcal{D}\left(\omega,\,k;\,\rm Ri\right)=\left|
\begin{array}{ll}
I_\nu(-k-\omega) & K_{\nu}(-k-\omega) \\
I_\nu(k-\omega)  & K_{\nu}(k-\omega)
\end{array}
\right|=0,
\label{eqn:desRel_komega}
\end{equation}
where $c \notin [-1,1]$. It is verified that the dispersion relation remains invariant under the transformation $\omega \rightarrow -\omega$. Thus at a fixed wavenumber ($k$), for each forward propagating (prograde) mode with frequency ($\omega$) there always exists a backward propagating (retrograd) mode with frequency ($-\omega$).

The linear system~\eqref{eqn:AB} has infinitely many solutions for $A$ and $B$ when the coefficients satisfy~\eqref{eqn:desRel_kc} or \eqref{eqn:desRel_komega}. From these infinite set of solutions we choose $A = K_{\nu}(k(-1-c))$ and $B = - I_\nu(k(-1-c))$. Finally substituting expressions of $A$ and $B$ into~\eqref{eqn:gen_sol_psi}, we get
\begin{equation}
\label{eq:psi_hat}
\hat{\psi}(z)=\sqrt{k(z-c)}\left[K_{\nu}(k(-1-c))I_\nu(k(z-c))-I_\nu(k(-1-c))K_{\nu}(k(z-c))\right].
\end{equation}

Under the long wavelength approximation, i.e.~$k=0$, the explicit expressions of the phase speed $c$ satisfying~\eqref{eqn:desRel_kc} and the eigenfunction $\hat{\psi}(z)$ are given in Ref.~\cite{eliassen1953two,davey_reid_1977}, which are as follows 
\begin{align}
    c_j = \coth\, {\frac{j\,\pi}{2\sqrt{\rm Ri - 0.25}}}, \qquad \mbox{and} \qquad j={1,2,3\cdots},
    \label{eq:eigval_k0}
\end{align}
\begin{align}
    \hat{\psi}_j(z)=(-1)^{j-1}\frac{(c_j+1)^{\frac{1}{2}}(c_j-z)^{\frac{1}{2}}}{\sqrt{\rm Ri - 0.25}}\sin\left\{{\sqrt{\rm Ri - 0.25}\log{\left(\frac{c_j-z}{c_j-1}\right)}}\right\}.
\end{align}
For $k\neq 0$, the explicit expressions of phase velocity and eigenfunctions are too cumbersome to obtain. 
From~\eqref{eq:eigval_k0}, we can see that for every $\rm Ri>0.25$ the sequence of phase speeds $c_j$ converges to $1$ as $j\rightarrow \infty$.

We shall solve the dispersion relation $\mathcal{D}(\omega,k; \rm Ri)=0$, as follows, (i) for fixed wavenumber ($k$) to obtain an infinite set of frequencies $\omega_1,\,\omega_2,\cdots$, and also (ii)  for fixed frequency ($\omega$) to obtain an infinite set of wavenumbers $k_1,\,k_2,\cdots$. 

\begin{table}[!htbp]
\centering
\begin{tabular}
{|>{$}c<{$}|>{$}c<{$}|>{$}c<{$} |>{$}c<{$} |>{$}c<{$} |>{$}c<{$} |>{$}c<{$} |>{$}c<{$} |>{$}c<{$} |>{$}c<{$}| >{$}c<{$}|}
\hline
\omega \to & 1.45442 & 2.0197 & 2.53705 & 3.04056 & 3.541166&1.09730 & 1.61782 & 2.12460 & 2.626178 & 3.12646 \\
\hline 
k_1 & 0.999998& 1.49994 & 2.0 & 2.5 & 3.0 & 0.721779 & 1.13742 &	1.59915	& 2.08802 & 2.58572\\ \hline 
k_2 & 1.34115& 1.89586&2.441 & 2.91411& 3.41465&1.0	&1.5	&2.0	&2.5&	3.0\\ \hline 
k_3&1.42731&1.99033&2.50723 & 3.01066& 3.51125&1.07379	&1.58973&	2.09508&	2.59633&	3.09657\\ \hline 
k_4&1.44799&2.01275&2.52999&3.03348 & 3.53409&1.09172	&1.61116&	2.11762&	2.61912&	3.11939\\ \hline 
k_5&1.4529&2.01805&2.53538&3.03888&3.53949&1.09598&	1.61625&	2.12295&	2.62451	&3.12479\\ \hline 
k_6&1.45406&2.01931&2.53666& 3.04016 &3.54107&1.09699&	1.61745&	2.12422&	2.62578&	3.12607\\ \hline 
k_7&1.45433&2.01961&2.53696& 3.04046&3.54114&1.09723&	1.61773	&2.12452&	2.62608	&3.12637\\ \hline
k_8&1.45442&2.01969&2.53704& 3.04054&3.541166&1.09729&	1.6178&	2.12459&	2.62616&	3.12645\\
\hline
\end{tabular}
\caption{\small{
The roots of the dispersion relation~\eqref{eqn:desRel_komega} for ${\rm Ri}=5$ showing the
presence of different modes of wavenumbers $k_1$--$k_8$ for  several frequencies.   
}}
\label{tab:w_k_Ri5}
\end{table}

For $\omega>k$, the wavenumber $k$ can be expressed as $k=\omega -\delta$, where the frequency $\omega$ is fixed and $\delta>0$ is an unknown. This further implies that solving $\mathcal{D}(\omega,\omega-\delta; \rm Ri )=0$ for $\delta$ is sufficient for finding the dispersion relation as a function of $\omega$. Note that, we can also solve the dispersion relation as a function of wavenumber in the same way. Similarly, under the condition $\omega<-k$, we can find $\omega$ and $k$, satisfying the dispersion relation. We use Mathematica software for solving the dispersion relation~\eqref{eqn:desRel_komega}. 
In addition, the linear stability problem is also solved numerically as a generalized eigenvalue value problem by using spectral collocation method. Table~\ref{tab:w_k_Ri5} summarizes the roots of the dispersion relation, which shows the presence of various waves of different wavenumbers at a fixed frequency.  
 

\begin{figure}[!htb]
\centering
\vspace{0.6em}
\includegraphics[scale=0.30]{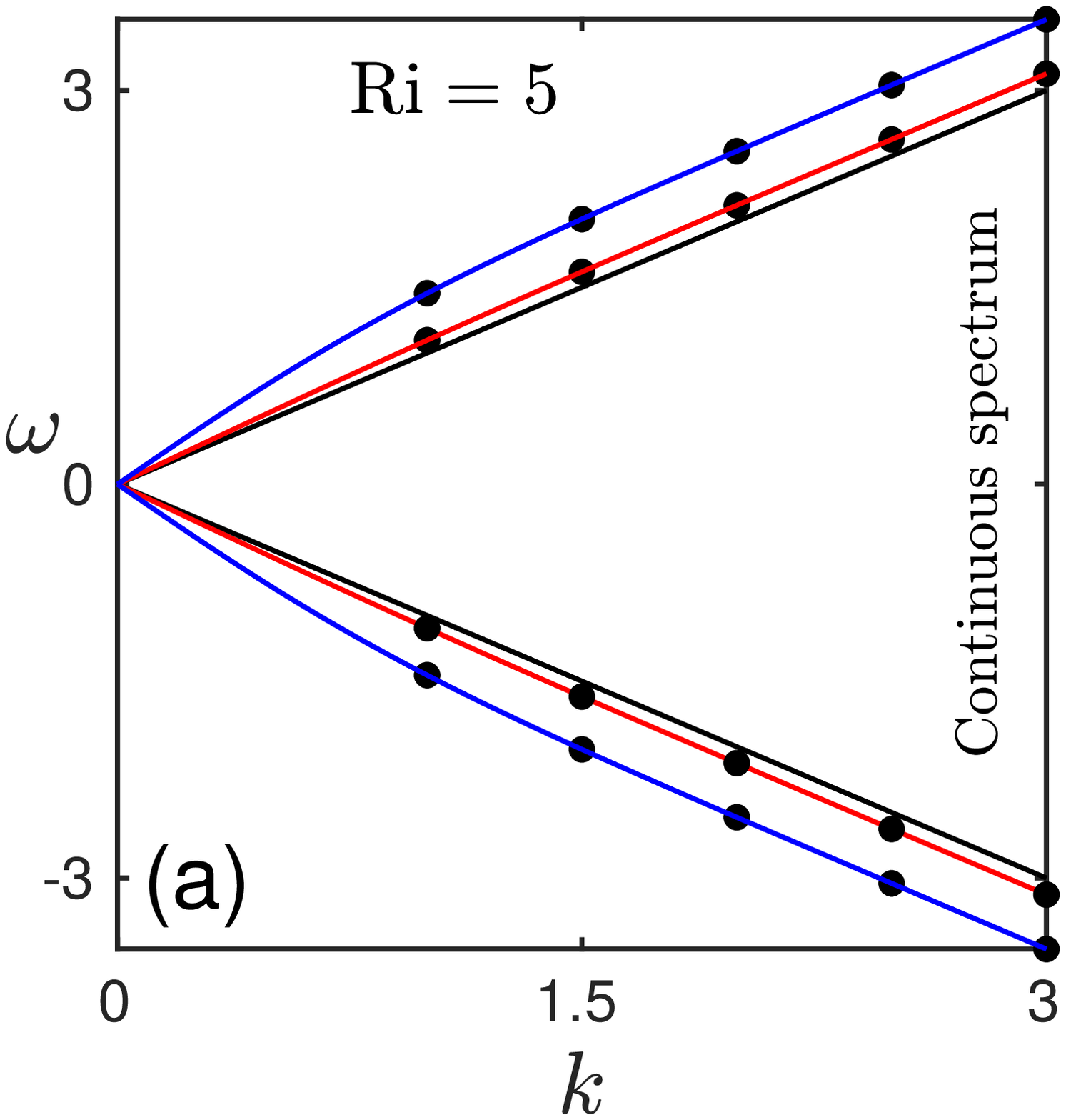}\qquad \qquad
 \includegraphics[scale=0.29]{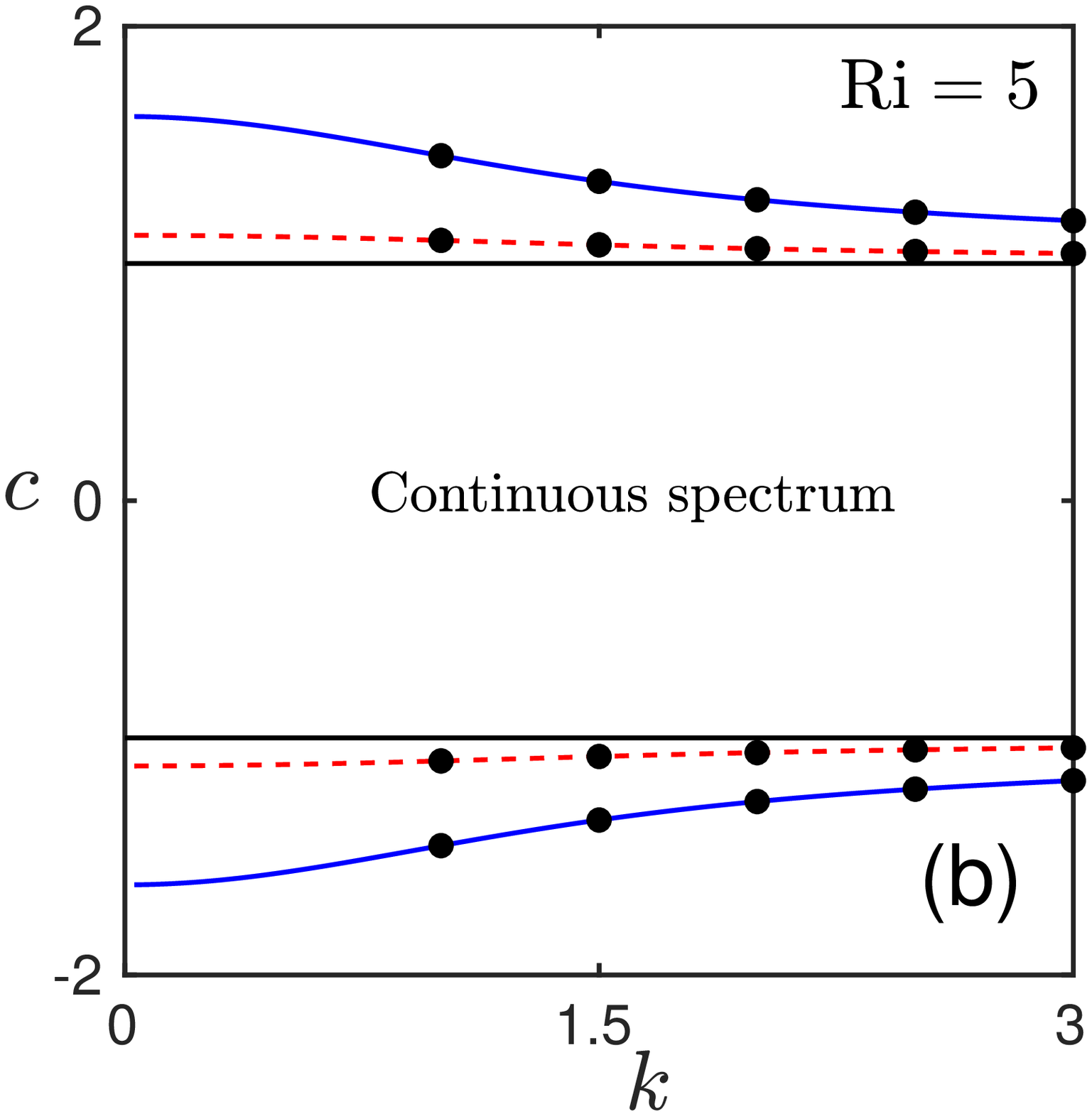}
\caption{\small{
Variation of the (a) frequency $\omega$ 
and (b) phase speed $c=\omega/k$ with the wavenumber for ${\rm Ri}=5$.  
}}
\label{fig:dis_rel}
\end{figure}

Figure~\ref{fig:dis_rel}(a) illustrates the variation of the frequencies of two prograde and two retrograde modes  with wavenumbers. In figure~\ref{fig:dis_rel}(a) circles are the $\omega$'s calculated from the dispersion relation~\eqref{eqn:desRel_komega} and blue (red) line denotes numerically computed frequencies of the first (second) prograde and the corresponding retrograde mode obtained from the eigenvalue problem~\eqref{eqn:gen_evp_implicitform}. It is clearly seen that (i) numerically computed eigenvalues are in excellent agreement with those of obtained from the dispersion relation, and (ii) the frequency increases (decreases) with increasing wavenumber for prograde (retrograde) modes and frequencies of different modes eventually converge to $\omega=k$ ($\omega=-k$), shown by thick solid line. The variation of corresponding phase speeds is shown in figure~\ref{fig:dis_rel}(b). Clearly, the phase speed of all types of modes approaches to $c=\pm 1$ with increasing $k$. As mentioned earlier in this section, in the present work we restrict ourselves to $c>1$ ($\omega>k$), i.e.~prograde modes.

\begin{figure}
\vspace{0.5em}
\centering
\includegraphics[scale=.3]{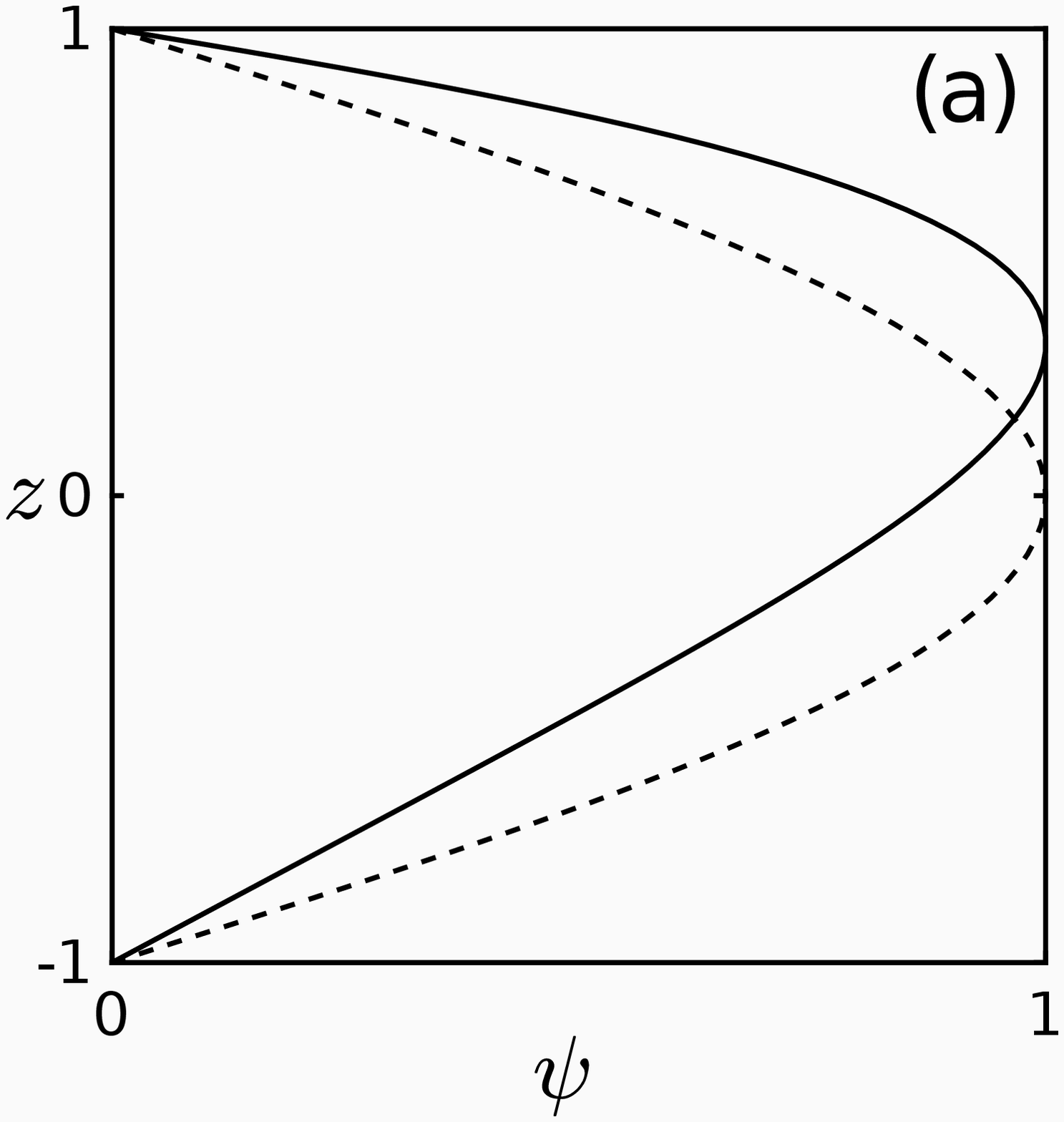}\quad
\includegraphics[scale=.3]{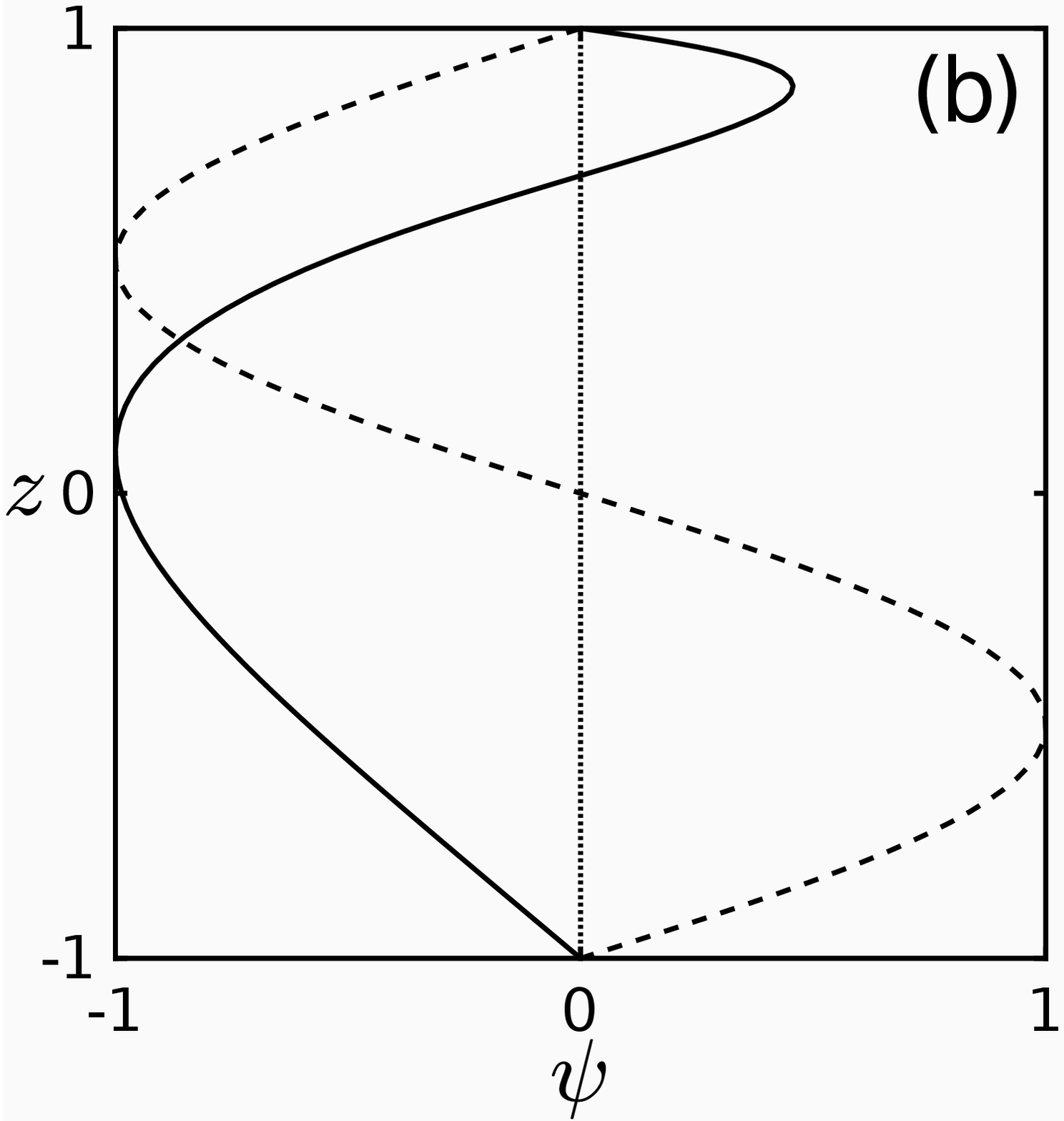}\quad
\includegraphics[scale=.3]{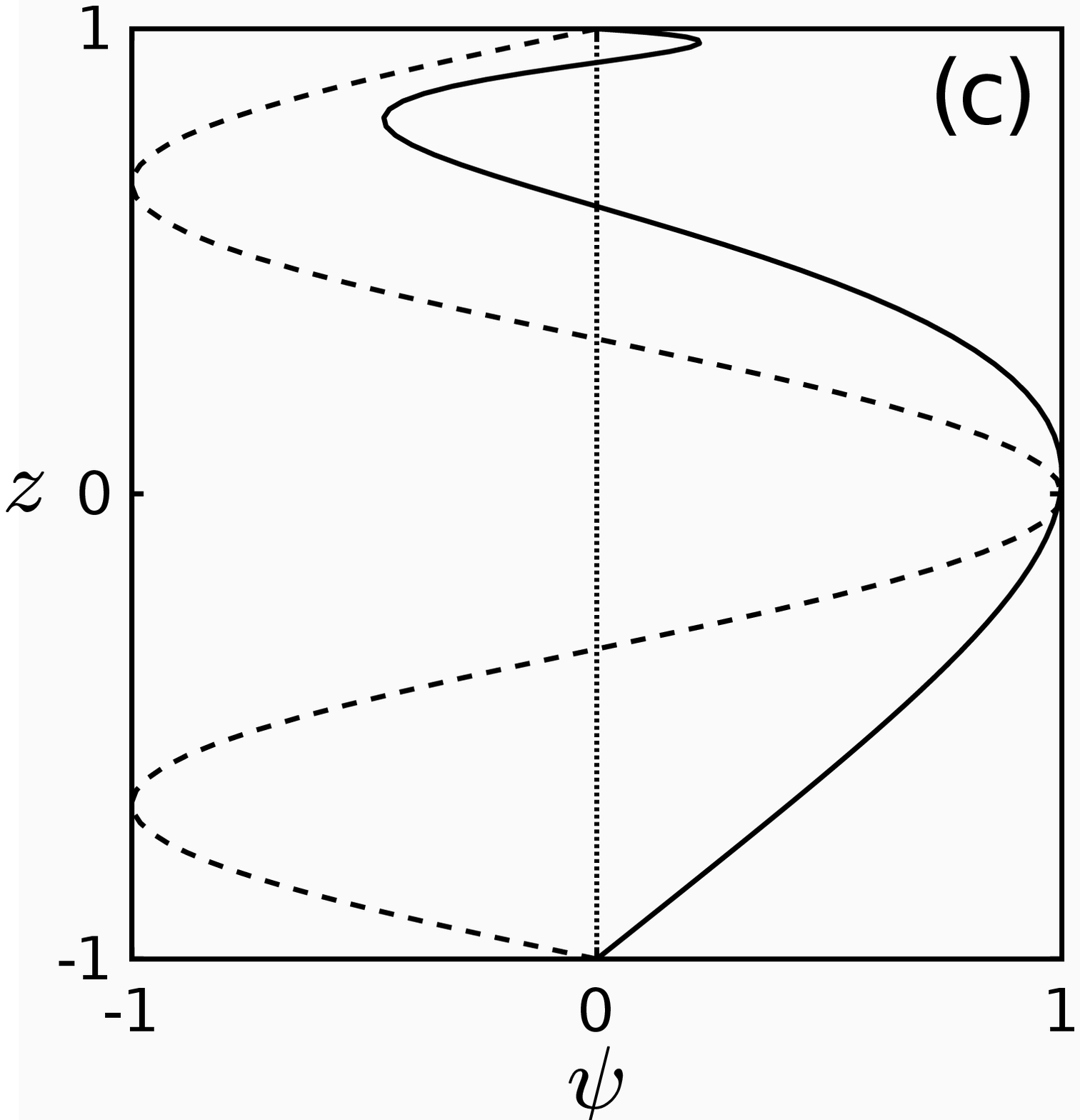}
\caption{\small{
Normalized eigenfunctions representing (a) mode~1, (b) mode~2 and (c) mode~3 for ${\rm Ri}=5$ and $k=1$.  
The solid and dashed line denotes the case with shear and without shear, respectively.    
}}
\label{fig:efunction}
\end{figure}

The variation of three eigenfunctions of~\eqref{eq:tg_linear} with background shear (solid line) and without background shear (dashed line) for $k=1$ and ${\rm Ri}=5$ is depicted in figure~\ref{fig:efunction}. It is worth noticing that in the absence of shear flow, the eigenmodes are nothing but the sine and cosine functions, which transform to the modified Bessel functions in the presence of uniform shear flow. The mode number of an eigenfunction is related to the number of zero crossing as well as number of extremum of the eigenfunction---if $n$ is the mode number of the eigenmode then the eigenmode becomes zero $n-1$ times and achieves extremum $n$ times between the domain $z\in (-1,1)$. Thus, the modes with mode number one, two and three cross the zero line (dotted line) zero, one and two times, respectively, and reach extremity one, two and three times, respectively, see figure~\ref{fig:efunction}. Clearly, the eigenfunctions are asymmetric in the presence of background shear flow.

\section{Nonlinear problem}
\label{sec:Nonlinear problem}
For examining nonlinear problem, we seek an approximate solution of~\eqref{eqn:dimlessrho} and~\eqref{eqn:tg_nonlinear} in the following expansion form,
\begin{equation}
\left(\hat{\psi},\,\hat{\rho}\right)=
\epsilon\,\left(\psi^{(1)},\,\rho^{(1)}\right)+\epsilon^2\,\left(\psi^{(2)},\,\rho^{(2)}\right)+\cdots,
\label{eqn:expansion}
\end{equation}
where $\epsilon$ is a small dimensionless parameter. In particular, we focus on the second-order solution $\left(\psi^{(2)},\,\rho^{(2)}\right)$. Substituting~\eqref{eqn:expansion} into~\eqref{eqn:dimlessrho} and~\eqref{eqn:tg_nonlinear} and equating the coefficients of like powers of $\epsilon$, we obtain a system of equations for $\left(\psi^{(n)},\, \rho^{(n)}\right)$ which are solved successively. At $\mathcal{O}(\epsilon)$, we obtain
the linear disturbance equations
\begin{equation}
\label{linear_eq}
\mathcal{L}\,\psi^{(1)} = 0\quad \mbox{and}\quad
\left( \partial_t + \bar{U}(z)\, \partial_x\right)\rho^{(1)} +
\frac{N^2}{N_0^2}\,\partial_x \psi^{(1)}=0,
\end{equation}
where $\mathcal{L} = \left( \partial_t + \bar{U}(z) \partial_x \right)^2\nabla^2  + {\rm Ri}\, \partial_{xx}$, where the subscripts $x$ and $t$ denote the respective derivatives. Similarly at $\mathcal{O}(\epsilon^2)$,
we arrive at the following nonlinear system,
\begin{subequations}
\begin{gather}
\mathcal{L}\,\psi^{(2)} = \left( \partial_t + \bar{U}(z) \partial_x \right)
J\left(\psi^{(1)},\,\nabla^2\psi^{(1)}\right)
+
{\rm Ri}_0 \,\partial_x J\left(\psi^{(1)},\rho^{(1)}\right),
\label{eqn:2ndorder_psi_a}
\\
\left( \partial_t + \bar{U}(z) \partial_x\right)\rho^{(2)} +
\frac{N^2}{N_0^2}\,\partial_x \psi^{(2)}
=J\left(\psi^{(1)},\,\rho^{(1)}\right).
\label{eqn:2ndorder_psi_b}
\end{gather}
\label{eqn:2ndorder_psi}
\end{subequations}

\subsection{First-order solution}
\label{subsec:First order solution}

The first-order equations~\eqref{linear_eq} are identical with~\eqref{eqn:gen_evp_implicitform} and~\eqref{eq:tg_linear} and thus, we follow the similar approach as discussed in \S\ref{sec:Linear stability theory}. Let $\left(\psi^{(1)},\rho^{(1)}\right)$ be in form of Fourier series which represents a superposition of waves of different wavenumbers
\begin{equation}
\psi^{(1)}= \sum_{m}{\psi}_m(z)e^{i\left(k_m\,x-\omega_m t\right)}+\mbox{c.c.}\quad \mbox{and}\quad  \rho^{(1)}= \sum_{m}{\rho}_m(z)e^{i
\left(k_m\,x-\omega_m t\right)}+\mbox{c.c.},
\label{eqn:sol_psi1rho1}
\end{equation}
where $\omega_m=k_m c_m$ with $m\geq1$ being a positive integer and $\mbox{c.c.}$ denotes the complex conjugate terms. Substituting~\eqref{eqn:sol_psi1rho1} into~\eqref{linear_eq}, we get 
\begin{equation}
L_m {\psi}_m(z)=0 \quad \mbox{and} \quad
{\rho}_m(z) = -\frac{N^2}{N_0^2} \frac{{\psi}_m(z)}{\left(\bar{U}(z) - \omega_m/k_m\right)},
\end{equation}
where
$L_m \equiv \mathcal{L}( \partial_t \rightarrow - i \omega_m, \partial_x \rightarrow i k_m, \partial_{xx} \rightarrow - k_m^2, \partial_z \rightarrow {\rm d}/{\rm d}z, \dots)$ is a complex differential operator. As detailed in \S\ref{sec:Linear stability theory}, the exact solution of the first-order problem is expressed in terms of the modified Bessel function as
\begin{align*}
{\psi}_m(z) &=\sqrt{p_m}\left[K_{\nu }\left(-k_m-\omega_m \right) I_{\nu }(p_m)-I_{\nu }\left(-k_m-\omega_m \right) K_{\nu }(p_m)\right],
\end{align*}
where $p_m=k_m z - \omega_m$ and $\psi_m(\pm1)=0$, see~\eqref{eq:psi_hat}. 

\subsection{Second-order solution and three-wave interactions}
\label{subsec:2ndsol}

Note that the right-hand side of the second-order system~\eqref{eqn:2ndorder_psi} is forced by the solution $\left(\psi^{(1)},\rho^{(1)}\right)$ of the first-order system. Owing to the forcing by primary (linear) waves, one can physically interpret the generation of higher order harmonics (waves which are forced by linear waves) in various ways, for instance, (i) two linear waves interact with each other, exchange energies, and give rise to the second harmonic and the correction of the mean term, and (ii) these three waves, two waves at the first-order and the second harmonic, further interact and produce the higher order harmonics, and so on. At resonance condition, three interacting waves affect each other significantly due to which solution of~\eqref{eqn:2ndorder_psi} produces large amplitude. These interacting waves form a resonance triad if the sum of their wavenumbers and frequencies satisfy the following conditions,
\begin{align}
k_1 + k_2 = k_3\quad \mbox{and}\quad\omega_1 + \omega_2 =\omega_3,
\label{eqn:triad_condition}
\end{align}
where $k_1$ and $k_2$ are the wavenumbers of two interacting waves that produce a wave of wavenumber $k_3$ with $\omega_1$, $\omega_2$ and $\omega_3$ being the corresponding frequencies. Here, the numbered subscript is used for the purpose of labeling interacting waves. 

In this paper we consider the interaction between two primary modes of wavenumbers $k_1=k_m$ and $k_2=k_n$ having the same frequency $\omega_1=\omega_2=\omega$. These two primary modes interact with each other and produce two types of secondary modes, one is the superharmonic mode  and another is the subharmonic mode. The superharmonic and subharmonic modes have the wavenumber and frequency pairs as $\left(k_m+k_n,2\omega\right)$ and $\left(k_m-k_n,0\right)$, respectively. The superharmonic mode form a resonant triad with two interacting primary modes when $2\omega$ is one of the eigenvalues of ~\eqref{eqn:gen_evp_implicitform} at wavenumber $k_m+k_n$, i.e.~$(k_m+k_n,2\omega)$ satisfies the dispersion relation $\mathcal{D}(2\omega,\,k_m+k_n;\, \rm Ri)=0$. Similarly, the subharmonic mode form a resonant triad with two interacting primary modes when $0$ is one of the eigenvalues of the linearized problem ~\eqref{eqn:gen_evp_implicitform} at wavenumber $k_m-k_n$, i.e.~$(k_m-k_n,0)$ satisfies the dispersion relation $\mathcal{D}(0,\,k_m-k_n;\, \rm Ri)=0$. In other words, under the resonance condition there exists a primary internal  mode with wavenumber and frequency pair as $\left(k_m+k_n,2\omega\right)$ or $\left(k_m-k_n,0\right)$ in the system.

Writing the right-hand side of~\eqref{eqn:2ndorder_psi_a} as a  combination of primary waves $\left(\psi^{(1)},\rho^{(1)}\right)$ of wavenumber $k_m$ and $k_n$,

\begin{align}
\sum_{m,n} \left[ A_{mn}(z) \,e^{i\,[(k_m+k_n)x -  2 \,\omega\, t]}
       + B_{mn}(z) \,e^{i\,(k_m-k_n)x} + \,\mbox{c.c}\right],
\label{eqn:N1N2_mn}    
\end{align}
where
\begin{subequations}
\begin{align}
  A_{mn}(z) = \left(k_m+k_n\right)&\left[ - \left(\bar{U}(z) - c_3\right)\,
\left( k_m \,{\psi}_m\,(D^2-k_n^2)\,D{\psi}_n-k_n \,D{\psi}_m\,(D^2-k_n^2){\psi}_n \right)\right .
\nonumber
\\
 &\left.+ {\rm Ri}_0 \,
\left(
-k_m\,{\psi}_m\,D{\rho_n}+ k_n \,{\rho}_n\,D{\psi}_m\right)\,\right],
\label{eqn:A_mn}
\\
 B_{mn}(z) = - (k_m-k_n)&\left[ \,\bar{U}(z)\, k_m\, {\psi}_m\,(D^2-k_n^2)\,D{\psi}_n + \bar{U}(z)\, k_n  \,D{\psi}_m\,(D^2-k_n^2){\psi}_n \right.
\nonumber\\
 &\left.+ {\rm Ri}_0\, \left(k_m\, {\psi}_m \, D{\rho}_n 
-  k_n\,{\rho}_n\,D{\psi}_m \right)\,\right]. 
\label{eqn:B_mn}
\end{align}
\label{eqn:AB_mn}
\end{subequations}
Here, the phase velocity of the superharmonic mode is defined as $c_3= 2 \omega/(k_m+k_n)$ and the subharmonic mode is independent of time. 

Note that, the right-hand side forcing~\eqref{eqn:N1N2_mn}--\eqref{eqn:AB_mn} arises due to the  quadratic interaction between various modes present at the leading order. Consequently, the second-order solution $\psi^{(2)}$ has the same form as the two terms of the right-hand side forcing of~\eqref{eqn:2ndorder_psi_a}, thus the expression for $\psi^{(2)}$ reads
\begin{align}
\psi^{(2)}(z)= \sum_{m,n} \left[h_{mn}(z) e^{i[(k_m+k_n)x -  2\omega t]}
       + g_{mn}(z) e^{i(k_m-k_n)x} + \mbox{c.c}. \right].
\label{eqn:phi2_sol}
\end{align}
Substituting~\eqref{eqn:phi2_sol} into~\eqref{eqn:2ndorder_psi_a} and equating the coefficients of $e^{i[(k_m+k_n)x -  2\omega t]}$ and $e^{i(k_m-k_n)x}$, we obtain the following governing equations for $\bar{h}_{mn}(z)$ and $\bar{g}_{mn}(z)$
\begin{align}
L_2^{+} \bar{h}_{mn}(z) &= 
\bar{A}_{mn}(z), 
\label{eqn:hmn}\\
L_2^{-}\, \bar{g}_{mn}(z) &= \bar{B}_{mn}(z) ,
\label{eqn:gmn}
\end{align}
where $\bar{A}_{mn} =A_{mn}+A_{nm}$ and $\bar{B}_{mn} =B_{mn}+B_{nm}$ with $A_{nm}$ and $B_{nm}$ being obtained by interchanging $m$ and $n$ indices in the expressions of $A_{mn}$ and $B_{mn}$, respectively; $L_2^+$ and $L_2^-$ are the linear differential operators defined as
\begin{align}
L_2^+ &\equiv  \mathcal{L}\left(\frac{\partial}{\partial x} \rightarrow i(k_m+k_n),\frac{\partial}{\partial z}  \rightarrow \frac{{\rm d}}{{\rm d}z}, \frac{\partial}{\partial t} \rightarrow -2i \omega, \dots \right) 
\nonumber
\\&= - \left(-2\omega+ \bar{U}(z)(k_m+k_n)\right)^2
(D^2-(k_m+k_n)^2)- {\rm Ri}\, (k_m+k_n)^2,
\label{eqn:L+}
\\
L_2^- &\equiv   \mathcal{L}\left(\frac{\partial}{\partial x} \rightarrow i(k_m-k_n),\frac{\partial}{\partial z}  \rightarrow \frac{{\rm d}}{{\rm d}z}, \frac{\partial}{\partial t} \rightarrow 0, \dots \right)
\nonumber\\
&=-\bar{U}(z)^2(k_m-k_n)^2(D^2-(k_m-k_n)^2)- {\rm Ri}\, (k_m-k_n)^2.
\label{eqn:L-}
\end{align}
Physically $\bar{h}_{mn}(z)$ and $\bar{g}_{mn}(z)$ represent the  spatial amplitudes  of the superharmonic and subharmonic waves, respectively. 

Note that, the superharmonic system~\eqref{eqn:hmn} is solvable iff
the corresponding homogeneous problem $L_2^+ \bar{h}_{mn}^h(z)=0$ possesses only trivial solution. Specifically, the system  $L_2^+ \bar{h}_{mn}^h(z)=0$ has a non-trivial solution when $2\omega$ is equal to one of the eigenvalues of the linear problem at wavenumber $k_m+k_n$. Moreover, the existence of a non-zero homogeneous solution of ~\eqref{eqn:hmn} is exactly the case of resonant triad interaction (RTI), see~\eqref{eqn:triad_condition}, because there exist primary internal modes with frequency and wavenumber pair as $(\omega, k_m)$, $(\omega, k_n)$ and $(2\omega, k_m+k_n)$. Owing to this fact, the divergence of $\bar{h}_{mn}(z)$ acts as an additional criterion for the existence of the RTI in the superharmonic case. 

Similarly, the solution of the subharmonic system ~\eqref{eqn:gmn}  diverges when the corresponding homogeneous system $L_2^- \bar{g}_{mn}^h(z)=0$ has the non-trivial solutions, that is the case when zero is one of the eigenvalues of the linear problem at wavenumber $k_m - k_n$. This is exactly the case of the RTI because there exist linear modes with frequency and wavenumber pairs as $(\omega, k_m)$, $(\omega, k_n)$ and $(0, k_m-k_n)$. Similar to $\bar{h}_{mn}(z)$, the divergence of $\bar{g}_{mn}(z)$ acts as an additional criteria for the possibility of the RTI in the subharmonic case. In this paper, we investigate the possibility of RTI among two primary internal modes at frequency $\omega$ and a superharmonic mode at frequency $2\omega$, thus we will focus only on the solution $\bar{h}_{mn}(z)$ of~\eqref{eqn:hmn}.

\subsection{Analytical solution of \texorpdfstring{$L_2^+ \bar{h}_{mn}(z) = \bar{A}_{mn}(z)$}{L2+ barhmn=barAmn} system}
\label{subsec:analytical_2ndsol}

Interestingly, the second-order  systems are analytically solvable using the method of variation of parameters. In the present work we are interested in the solution $\bar{h}_{mn}(z)$ of~\eqref{eqn:hmn}. Exploiting the method of variation of parameters, we find the particular solution of~\eqref{eqn:hmn} which enables us to find the general solution of it.

Let the general solution of the superharmonic system be
\begin{equation}
\bar{h}_{mn}(z)=C_1 \,f_1(z)+C_2\, f_2(z)+ f_1(z)\,u_1(z)+f_2(z)\,u_2(z),
\label{eqn:sol_hmn}
\end{equation}
where $C_1$ and $C_2$ are two arbitrary constants to be determined using the homogeneous boundary conditions at $z=\pm1$.
$\left\{f_1(z), f_2(z)\right\}$ forms a fundamental set of solutions of the corresponding homogeneous problem of~\eqref{eqn:hmn} which is expressed in terms of the modified Bessel function of complex order $\nu=\sqrt{0.25 - {\rm Ri}}$ as
\begin{align}
\left.
\begin{aligned}
f_1(z)&=\sqrt{(k_m+k_n)\,z-2\,\omega}\, I_\nu\left[  (k_m+k_n)z-2\,\omega) \right], 
\\
f_2(z)&= \sqrt{(k_m+k_n)\,z-2\,\omega}\, K_{\nu}\left[ (k_m+k_n)z-2\,\omega)\right],
\end{aligned}
\right\}
\label{eqn:f1f2}
\end{align}
and the set $\{u_1(z), u_2(z)\}$ has the integral form
\begin{align}
\left.
\begin{aligned}
&u_1(z)= - \int_{-1}^{z}
\frac{f_2(\zeta) \,\bar{A}_{mn}(\zeta)}{\left[2\,\omega- (k_m+k_n)\,\zeta\,\right]^2\, W(\zeta)}
\,{\rm d}\zeta, 
\\
&u_2(z)=  \int_{-1}^{z}
\frac{f_1(\zeta) \, \bar{A}_{mn}(\zeta)}{\left[2\,\omega- (k_m+k_n)\,\zeta\,\right]^2\, W(\zeta)}
\,{\rm d}\zeta,
\end{aligned}
\right\}
\label{eqn:u1u2}
\end{align}
where $u_1(-1)=u_2(-1) =0$ and $W(\zeta)=-(k_m+k_n)$ is the Wronskian of $f_1$ and $f_2$. In order to find the arbitrary constants, $C_1$ and $C_2$, we apply the Dirichlet boundary conditions $\bar{h}_{mn}(\mp 1)=0$, which give
\begin{align}
\left.
\begin{aligned}
C_1 \,f_1(-1)+C_2 \, f_2(-1) &= 0,
 \\
C_1 \, f_1(1)+C_2 \, f_2(1)+ f_1(1) \, u_1(1)+f_2(1) \, u_2(1)&=0. 
\end{aligned}
\right\}
\label{eqn:eqnsc1c2}
\end{align}
Solving the linear system~\eqref{eqn:eqnsc1c2}, we obtain
\begin{align}
C_1= - C_2 \, \frac{f_2(-1)}{f_1(-1)} \quad\mbox{and} \quad
C_2 = -\frac{f_1(-1)\,\left[\,f_1(1)\,u_1(1)+f_2(1)\,u_2(1)\,\right]}{-f_2(-1)\,f_1(1)+f_2(1)\,f_1(-1)}.
\label{eqn:C1C2}
\end{align}
Finally, substituting $C_1$ and $C_2$ from~\eqref{eqn:C1C2} into~\eqref{eqn:sol_hmn} we obtain the analytical solution,
\begin{align}
\bar{h}_{mn}(z)=& \left(\,f_1(1)\,u_1(1)+f_2(1)\,u_2(1)\,\right) \, \left( f_2(-1) \,f_1(z) - f_1(-1)\,f_2(z) \right)\nonumber\\ 
&+\left(\,f_2(-1)\,f_1(1)+f_2(1)\,f_1(-1)\,\right)\,\left(\,f_1(z)\,u_1(z)+f_2(z)\,u_2(z)\,\right),
\label{eqn:hmn_sol_analytical}
\end{align}
where $f_1$, $f_2$ and $u_1$, $u_2$ are given by~\eqref{eqn:f1f2} and~\eqref{eqn:u1u2}, respectively. The analytical solution presented in this section is obtained with the help of computer algebra software Mathematica. In addition, the second order solution is also computed numerically by solving directly~\eqref{eqn:hmn} using spectral collocation method in MATLAB.          

\begin{figure}[!htbp]
\vspace{0.5em}
\centering
\includegraphics[scale=.4]{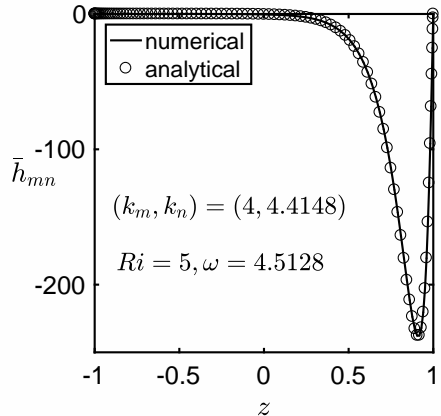}
\caption{
\small{Comparison of the solution $\bar{h}_{mn}(z)$ of the superharmonic system~\eqref{eqn:hmn}.
Solid line and circles denote the numerical and analytical solutions, respectively. 
}}
\label{fig:hmn_compare}
\end{figure}

Figure~\ref{fig:hmn_compare} depicts a comparison of analytical (circles) and numerical (solid line) solutions for ${\rm Ri}=5$, $\omega=4.5128$ and for two values of the wavenumbers $k_m$ and $k_n$ as shown in the panel. It is clearly seen from figure~\ref{fig:hmn_compare} that the analytical solution is in excellent agreement with those of numerically calculated ones. Henceforth, we shall use the analytical solution for most of the results except for the three-dimensional visualization results.

\section{Results}
\label{sec:Results}

As described above, the objective of the present paper is to determine the possibility of exact or near RTI among two primary modes at frequency $\omega$ and a superharmonic mode  at frequency $2\omega$ in a stably stratified uniform shear flow by analyzing interactions among the primary modes of varying frequencies. For simplicity, we consider four different primary modes at a fixed frequency $\omega$ and label their wavenumbers as $k_1$, $k_2$, $k_3$ and $k_4$. Here we shall investigate all possible interactions among any two of these modes of wavenumber $k_m$ and $k_n$ [$1\leq m\leq n\leq 4$], which form RTI with the superharmonic mode generated by them.

\subsection{Mode search method}

We shall now illustrate the mode search method to be adopted here for probing RTI.

\begin{enumerate}
  \item [STEP 1:] Generate $n_x\, \times \, n_y$ grid points on $(\omega, \rm Ri)$-plane and follow Step~2 to Step~8 for each grid point.

\item [STEP 2:]
Generate:
   \begin{enumerate}
       \item  $K_\omega=\{\,k\,\mid\,\mathcal{D}(\omega,k;\rm Ri)=0\,\} $.
       \item  $K_{2\omega}=\{\,k\,\mid\,\mathcal{D}(2\omega,k;\rm Ri)=0\,\} $.
   \end{enumerate}   
   
\item [STEP 3:] Arrange $K_\omega$ set in the ascending order and 
label the elements as $k_1,\, k_2, \, k_3, \, k_4, \cdots$.

\item [STEP 4:]  Calculate 
$k_m+k_n$ for $k_m$, $k_n \in K_\omega$, $1 \leq m \leq n \leq 4$.

        \item [STEP 5:]Generate: 
    $\Omega_{mn} = \{\, \omega \, |  \, \mathcal{D}(\omega,k_m+k_n;\rm Ri)=0 \,\}$ for $1 \leq m \leq n \leq 4$.

         \item [STEP 6:] Corresponding to each $(m,n)$-pair ($1 \leq m \leq n \leq 4$), define following sets:
    \begin{enumerate}
        \item $\Pi_\omega:=\{\, |2\omega-\omega_r| \, \mid \, \omega_r \in \Omega_{mn}\,\}$.
        \item $\Pi_k := \{ \, |k_m+k_n-k_r^{2\omega}|\, \mid \, k_r^{2\omega} \in K_{2\omega}\,\}$.
    \end{enumerate}
    
    \item [STEP 7:]
   For each mode pair $(m,n)$ ($1 \leq m \leq n \leq 4$), calculate minimum of $\Pi_\omega$ and $\Pi_k$
        \begin{enumerate}
        \item  $\Delta_\omega = \min \Pi_\omega$ and the corresponding
    $\omega_r$.
    \item  $\Delta_k = \min \Pi_k$ and the corresponding
    $k_r^{2\omega}$.
    \end{enumerate}

   \item [STEP 8:] 
    Determine the type of the resonant triad:
          \begin{enumerate}
      \item
Near RTI occurs if $\mathcal{O}(\Delta_\omega)$ or $\mathcal{O}( \Delta_k) \sim$ 
$10^{-4}$--$10^{-6}$.
\item
Exact RTI occurs if $\mathcal{O}( \Delta_\omega)$ or $\mathcal{O}( \Delta_k) \leq$ $10^{-7}$.
 \end{enumerate}
  \end{enumerate}

Without loss of generality, we assume that $n_x=50$, $n_y=80$, $\omega\in[0.01, 5]$,
and ${\rm Ri}\in[2,10]$ which implies that the effect of buoyancy is much more than the effect of velocity shear.

\subsection{Interactions among different modes \texorpdfstring{$m\neq n$}{m neq n}}
\label{subsec:m_notEQ_n}

\begin{table}
\centering
\scalebox{0.73}{
\begin{tabular}{
>{$}c<{$} >{$}c<{$}>{$}c<{$} >{$}c<{$} >{$}c<{$} >{$}c<{$} >{$}c<{$} >{$}c<{$} >{$}c<{$}}
{\rm Ri} & \omega & k_m& k_n &\omega_r & k_r^{2\omega}&\Delta_k&\Delta_\omega& \max|\bar{h}_{mn}|\\
\hline
\multicolumn{9}{c}{$(m, n)=(1, 2)$}\\ \hline
\rowcolor[gray]{0.8}
7.0&4.21&3.4327&4.2098 &8.4199& 7.64265 &1\times10^{-4}& 1.1151\times10^{-4} &3.6693\times10^4\\
7.0&4.31& 3.5327&4.3098 &8.6199& 7.84265&1\times10^{-4}& 1.2691\times10^{-4}&3.4419\times10^4\\
5.5&4.21&3.6080&4.2098 &8.4198& 7.81803 &2\times10^{-4}&1.4440\times10^{-4}&1.5948\times10^{4}\\
5.5&4.31&3.7080&4.3098 &8.6198& 8.01803&2\times10^{-4}&1.4890\times10^{-4}&1.6888\times10^4\\
5.5&4.41&3.8080&4.4098  &8.8198& 8.21803&2\times10^{-4}&1.5200\times10^{-4}&1.7954\times10^4\\ 
\hline
\multicolumn{9}{c}{$(m, n)=(1, 3)$}\\ \hline
\rowcolor[gray]{0.8}
 8.2&4.51&3.599&4.5100 &9.02000& 8.10899&5\times10^{-6}&9.2455\times10^{-7}&5.6420\times10^6
\\
7.0& 4.21& 3.4327&4.2100 &8.42000& 7.64265&4\times10^{-5}&1.1922\times10^{-6}&3.4336\times10^6\\
8.2& 4.61&3.6990&4.6100 &9.21998& 8.30899&5\times10^{-6}&1.2925\times10^{-5}&4.3073\times10^5\\
7.0&4.31&3.5327&4.3100 &8.61998& 7.84265&4\times10^{-5}&1.4208\times10^{-5}&3.0954\times10^5\\
8.2& 4.71&3.7990&4.7100 &9.41998& 8.50899 &5\times10^{-5}& 2.1125\times10^{-5}&2.7950\times10^5\\
\hline
\multicolumn{9}{c}{(m, n)=(1, 4)}\\ \hline
\rowcolor[gray]{0.8}
7.3& 4.21&3.3988&4.2099 &8.42& 7.60872&7\times10^{-8}&5.2173\times10^{-8}&8.0236\times10^7
\\
7.9&4.51&3.6319&4.5100 &9.02& 8.14189&1\times10^{-7}& 1.2038\times10^{-7}&4.3527\times10^7\\
8.1& 4.81&3.9099&4.8100 &9.62& 8.71992 &2\times10^{-5}&3.2267\times10^{-7}&1.9454\times10^7\\
5.5&4.31&3.7080&4.3100 &8.62& 8.01803&4\times10^{-7}&4.0420\times10^{-7}&6.2365\times10^6\\
7.3&4.51&3.6987&4.5100 &9.02& 8.20872&1\times10^{-6}&1.1522\times10^{-6}&4.4685\times10^6\\
 \hline
\multicolumn{9}{c}{(m, n)=(2, 3)}\\ \hline \rowcolor[gray]{0.8}
6.7&4.21&4.2048&4.2096 &8.41998& 8.41482&4\times10^{-4}&1.1682\times10^{-5}&4.2440\times10^6\\
2.0& 3.21&3.2086&3.2099 &6.42001& 6.40547&1\times10^{-4}&1.3219\times10^{-5}&3.5347\times10^6\\
6.7&4.31&4.3048& 4.3096 &8.62002& 8.61482&4\times10^{-4}&2.2322\times10^{-5}&1.8541\times10^6\\
6.4&1.21&1.0594&1.1663 &2.42004& 2.22566&4\times10^{-5}&3.6750\times10^{-5}&1.5914\times10^5\\
2.7&3.81&3.8055&3.8099 &7.61996& 7.61554&1\times10^{-4}&3.6815\times10^{-5}&5.0191\times10^6\\
 \hline
\multicolumn{9}{c}{(m, n)=(2, 4)}\\ \hline
\rowcolor[gray]{0.8}
2.2&1.81&1.7910&1.8098 & 3.62&3.60081&1\times10^{-5}&7.9970\times10^{-7}&2.9213\times10^7\\
3.7&1.71&1.6431&1.7077 &3.42& 3.35087&7\times10^{-5}&2.9214\times10^{-6}&1.0153\times10^6\\
5.1&1.71&1.5860&1.7028 &3.42001& 3.28873&7\times10^{-5}&6.7562\times10^{-6}&1.0676\times10^5\\
3.6&3.41&3.3449&3.4100 &6.81999& 6.75485&5\times10^{-5}&7.7865\times10^{-6}&1.1579\times10^6\\
3.6&3.51&3.4449&3.5100 &7.01999& 6.95485&5\times10^{-5}&9.1866\times10^{-6}&1.0185\times10^6\\
 \hline
\multicolumn{9}{c}{(m, n)=(3, 4)}\\ \hline
\rowcolor[gray]{0.8}
2.0&2.61&2.6086&2.6099 &5.22& 5.21865&1\times10^{-4}& 6.4018\times10^{-8}&4.4221\times10^8\\
2.0&2.71&2.7086&2.7099 &5.42& 5.41865&2\times10^{-4}&  4.5595\times10^{-6}&6.9869\times10^6\\
2.1&3.21&3.2083&3.2098 &6.41999& 6.41833&2\times10^{-4}&4.7191\times10^{-6}&8.2771\times10^6\\
2.0&2.51&2.5086&2.5099 &5.01999& 5.01865&1\times10^{-4}&5.1032\times10^{-6}&4.8302\times10^6\\
2.0&2.81&2.8086&2.8099 &5.62001& 5.61865&1\times10^{-4}& 8.1396\times10^{-6}&4.3229\times10^6\\
\hline
\end{tabular}}
\caption{\small{ Richardson number, frequency, wavenumbers associated with resonance. For each $(m,n)$-pair, this table shows some parameters for near or exact RTI predicted by $\Delta_\omega$ or $\Delta_k$. At exact resonance $\mathcal{O}\left(\Delta_\omega\right)$ or $\mathcal{O}\left(\Delta_k\right)\leq 10^{-7}$ and at near resonance $\mathcal{O}(\Delta_\omega)$ or $\mathcal{O}( \Delta_k) \sim$
$10^{-4}$--$10^{-6}$.}}
\label{tab:resonance_table1}
\end{table}

For different Richardson number ($\rm Ri$), obtained by varying the stratification, three waves satisfying closely the resonant conditions~\eqref{eqn:triad_condition} are identified and the results are summarized in table~\ref{tab:resonance_table1}. More precisely, at various ${\rm Ri}$ ($1^\mathrm{st}$ column) table~\ref{tab:resonance_table1} displays the frequency $\omega$ ($2^\mathrm{nd}$ column) with the corresponding wavenumbers $k_m$ and $k_n$ ($3^\mathrm{rd}$ and $4^\mathrm{th}$ columns) of two primary modes. As discussed above, at frequency $\omega$ these two interacting modes may or may not form a resonant triad with the third mode. Table~\ref{tab:resonance_table1} also exhibits the resonating frequency $\omega_r$ ($5^\mathrm{th}$ column) and the  resonating wavenumber $k_{r}^{2\omega}$ ($6^\mathrm{th}$ column), which form the near or exact resonant triad with the primary waves of wavenumbers $k_m$ and $k_n$.
The parameters~$\Delta_k$ ($7^\mathrm{th}$ column) and $\Delta_\omega$ ($8^\mathrm{th}$ column) classify the resonant triads as near or exact by measuring the accuracy in the resonance condition~\eqref{eqn:triad_condition}.
The last column of table~\ref{tab:resonance_table1} shows the maximum of the second-order solution, i.e.~$\max|\bar{h}_{mn}|$. Note that at the resonating frequency $\omega_r$, $\max|\bar{h}_{mn}|$ has a large value as compared to the neighboring non-resonating frequencies. 

Except for the case of $(m,n)=(1,2)$ and $(2,3)$, it is found that in all mode pairs $(m,n)$ the resonance conditions~\eqref{eqn:triad_condition} are exactly satisfied for some ${\rm Ri}$ as $\Delta_\omega$ or $\Delta_k$ $\leq \mathcal{O}(10^{-7})$, see the table. Nevertheless, it can be seen from table~\ref{tab:resonance_table1} that the near RTI occur for mode numbers $(m,n)=(1,2)$ and $(2,3)$ for some  $\omega$ and ${\rm Ri}$ values considered in this paper.

\subsubsection{Graphical way to analyse the evidence of a resonance triad}

We shall now analyse graphically the evidence of near or exact RTI to determine how accurately the resonance conditions are satisfied. Particularly, we seek the existence of a primary mode with wavenumber $k_m + k_n$ and frequency $2 \omega$---such a wave, if exists, forms a resonance triad with the two other primary modes of frequency and wavenumber pairs as ($\omega,k_m$) and ($\omega,k_n$), see~\S\ref{subsec:2ndsol}.    

\begin{figure}
\centering
\includegraphics[scale=0.26]{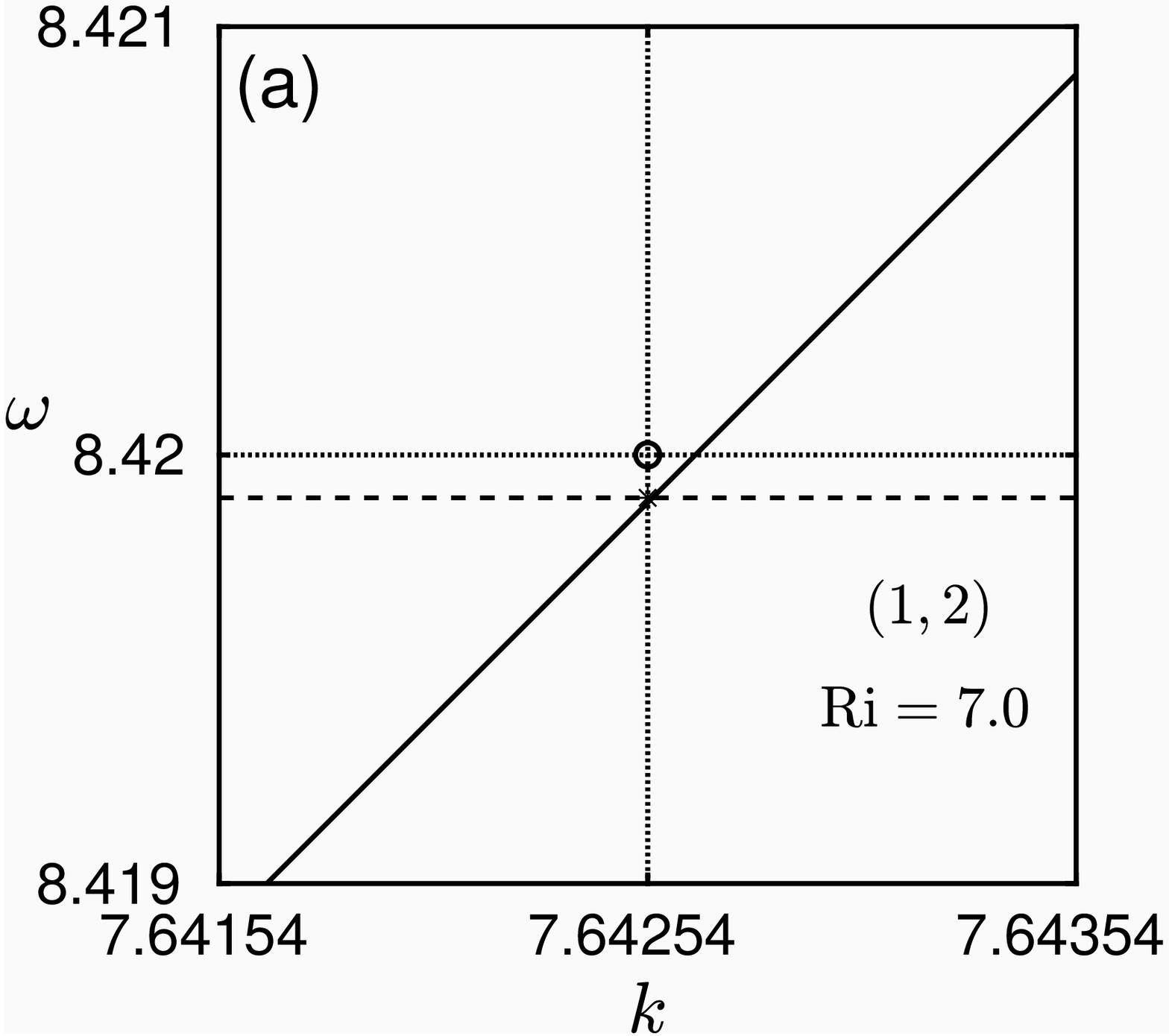}\qquad
\includegraphics[scale=0.24]{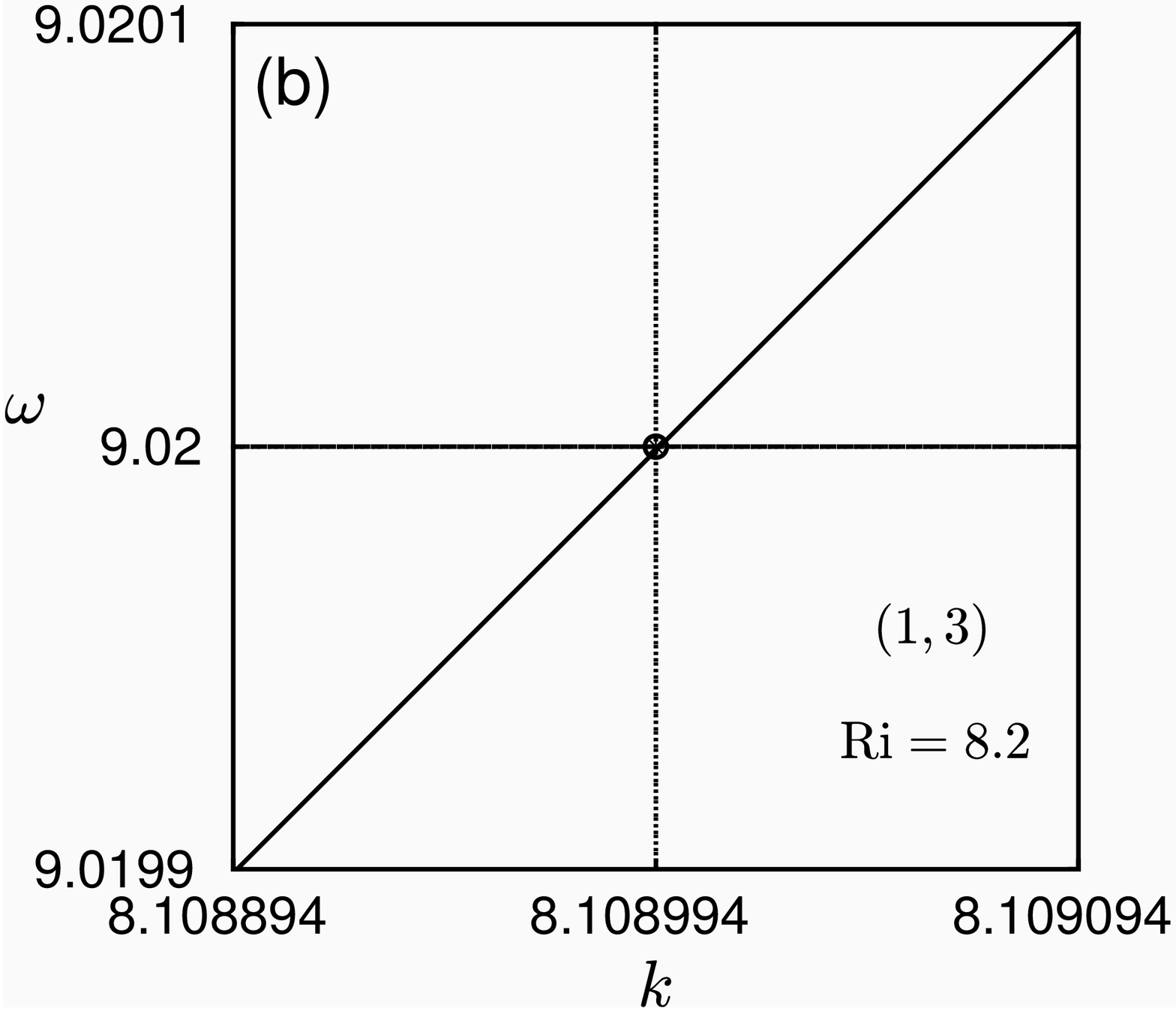}\qquad
\includegraphics[scale=0.24]{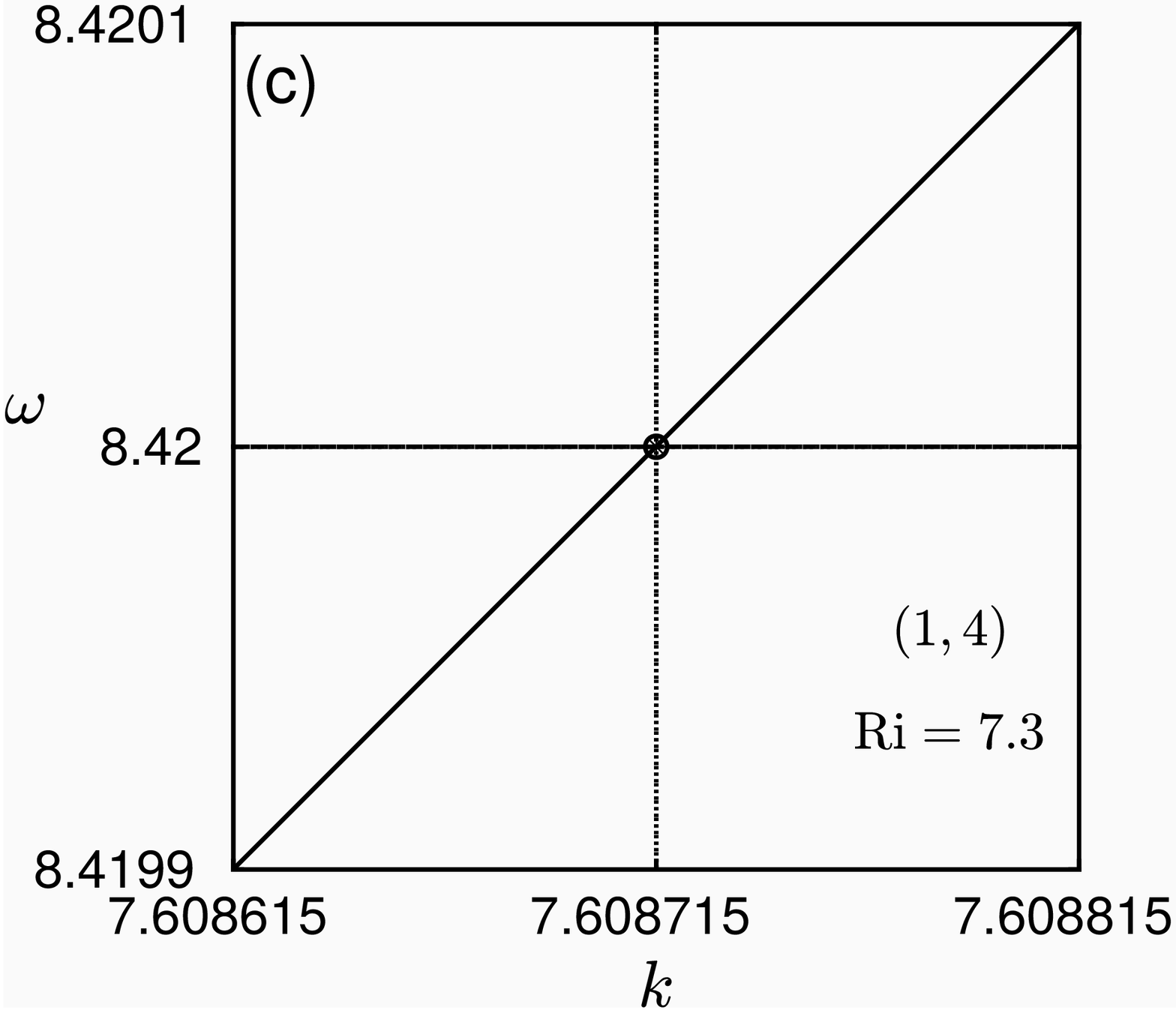}\\[1em]
\includegraphics[scale=0.24]{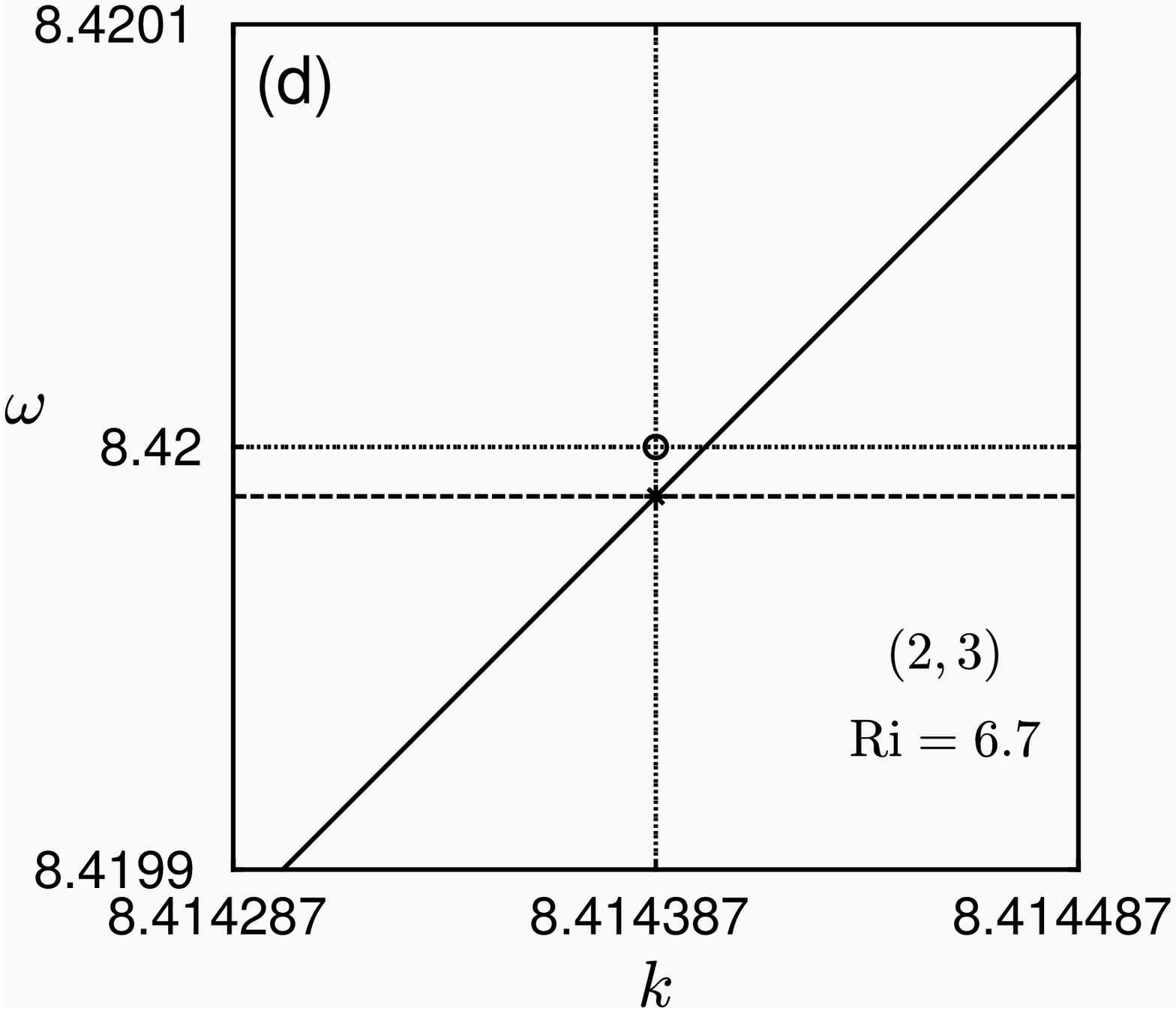}\qquad
\includegraphics[scale=0.24]{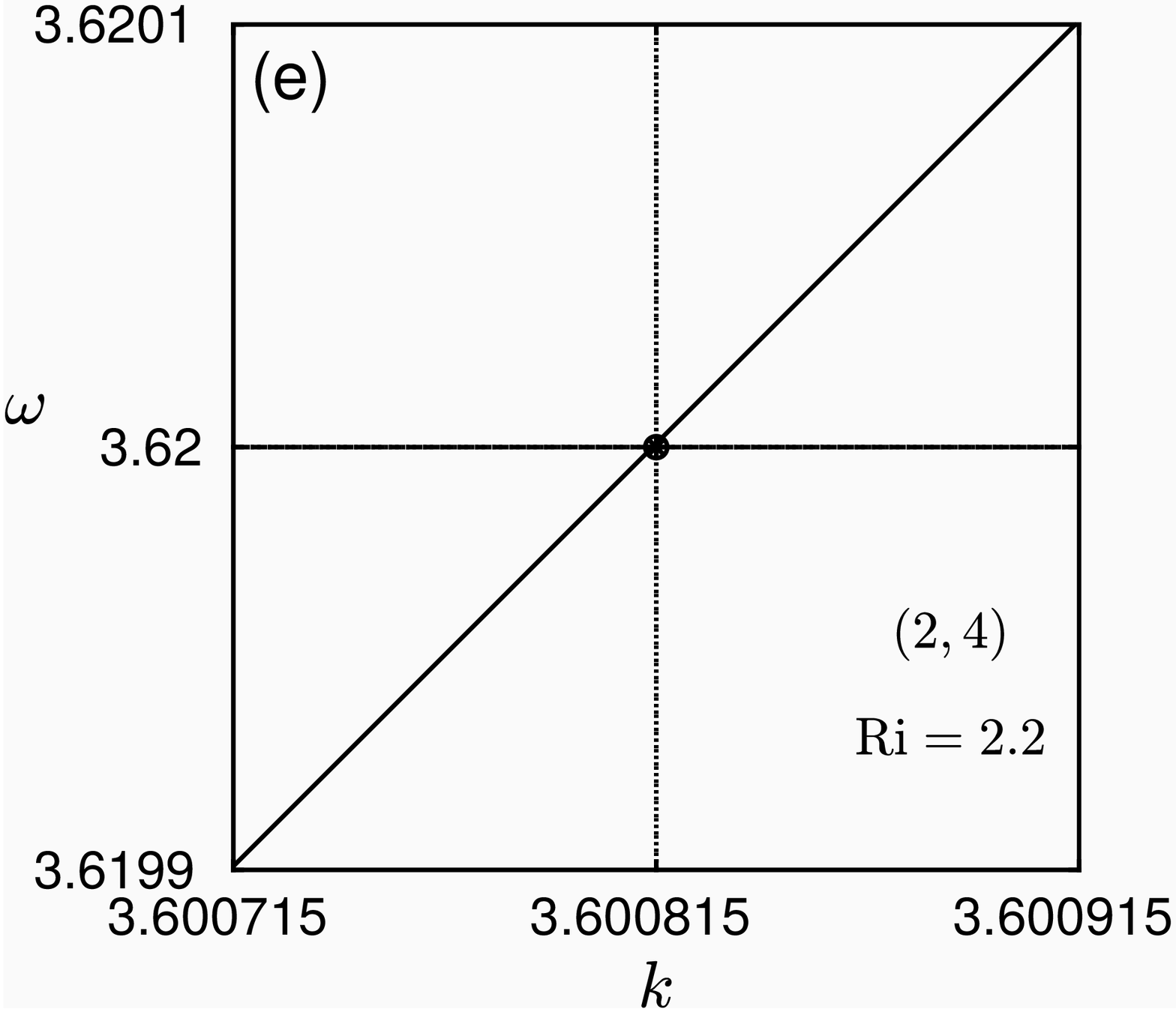}\qquad
\includegraphics[scale=0.24]{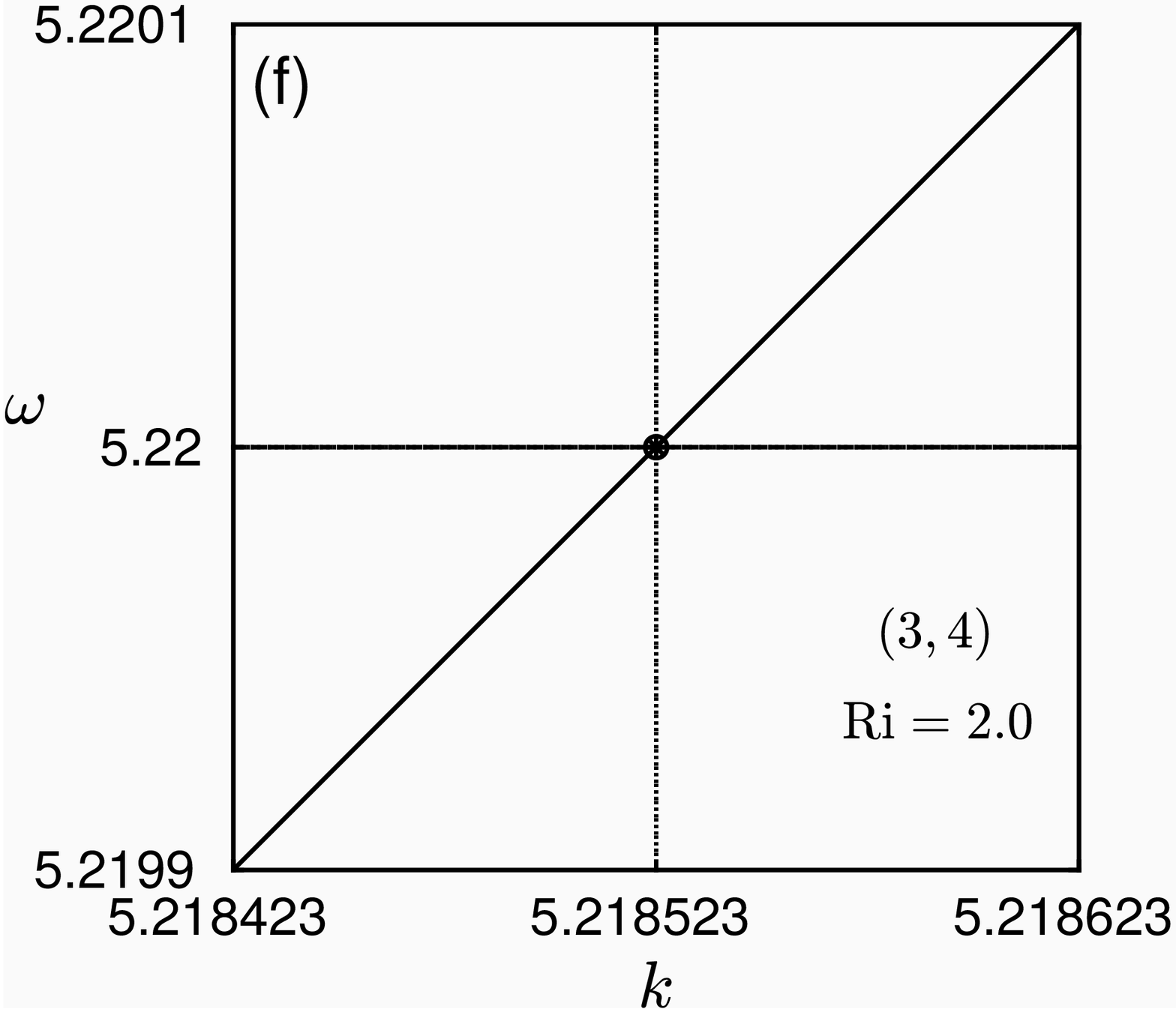}
\caption{\small{ Dispersion relation showing the evidence of resonance. Horizontal dashed and dotted lines represent $\omega_r$ (listed in 5th column of table~\ref{tab:resonance_table1}) and double of the frequency ($2\omega$) of the primary modes of wavenumbers $k_m$ and $k_n$ (listed in second column of table~\ref{tab:resonance_table1}), respectively. Dotted vertical line denotes $k=k_m+k_n$. Circle and star denote two points $(k_m+k_n,2\omega)$ and $(k_m+k_n,\omega_r)$, respectively. The vertical distance between these two points represents $\Delta_\omega$, which is zero for all $(m,n)$ pairs (panels b--f) except $(1,2)$ (panel a) and $(2,3)$ (panel d).  }
}
\label{2d plots}
\end{figure}

For each mode pair $(m,n)$, we choose ${\rm Ri}$ and frequency $\omega$ from the shaded row of table~\ref{tab:resonance_table1}. For this chosen value of ${\rm Ri}$, the dispersion curve (solid line) is depicted in figure~\ref{2d plots}. For the sake of comparison two points $(k_m+k_n,2\omega)$ (denoted by circle) and $(k_m+k_n,\omega_r)$ (denoted by star) are displayed in each panel, where $k_m$ and $k_n$ are the wavenumbers of two interacting primary waves and $\omega_r$ is the frequency of the superharmonic mode corresponding to the shaded rows of table~\ref{tab:resonance_table1}. Recall that the frequency and wavenumber pair ($\omega_r,\, k_m+k_n$) associated with the resonating mode satisfies the dispersion relation, thus the point ($k_m+k_n,\,\omega_r$), shown by star symbol, always lies onto the dispersion curve (solid lines in figures~\ref{2d plots}(a)--\ref{2d plots}(f)). On the other hand, in the situation when the resonance conditions are exactly satisfied~\eqref{eqn:triad_condition}, the point $(k_m+k_n,2\omega)$, shown by circle, also lies onto the dispersion curve and coincides with the point ($k_m+k_n,\omega_r$) leading to $\Delta_\omega=0$.

By observing figures~\ref{2d plots}(a)--\ref{2d plots}(f), we see that the point $(k_m+k_n,2\omega)$ lies exactly on the dispersion curve (solid line) for all mode pairs $(m,n)$ except $(m,n)=(1,2), (2,3)$. Consequently, a linear wave with wavenumber and frequency pair as $(k_m+k_n, 2\omega)$ exists for  mode pairs $(m,n)\neq (1,2),(2,3)$ and hence giving rise to the exact RTI among all mode pairs except $(1,2)$ and $(2,3)$. For $(m,n) = (1,2)$ and $(2,3)$, we observe that there is a small frequency mismatch ($\Delta_\omega$) in the temporal resonant condition. Thus, in these cases we define the interaction  as near RTI.

\subsubsection{Divergence surface for \texorpdfstring{$m\neq n$}{m neq n}}

\begin{figure}
\centering
\subfigure[]{
\includegraphics[width=6.5cm,height=4.0cm]{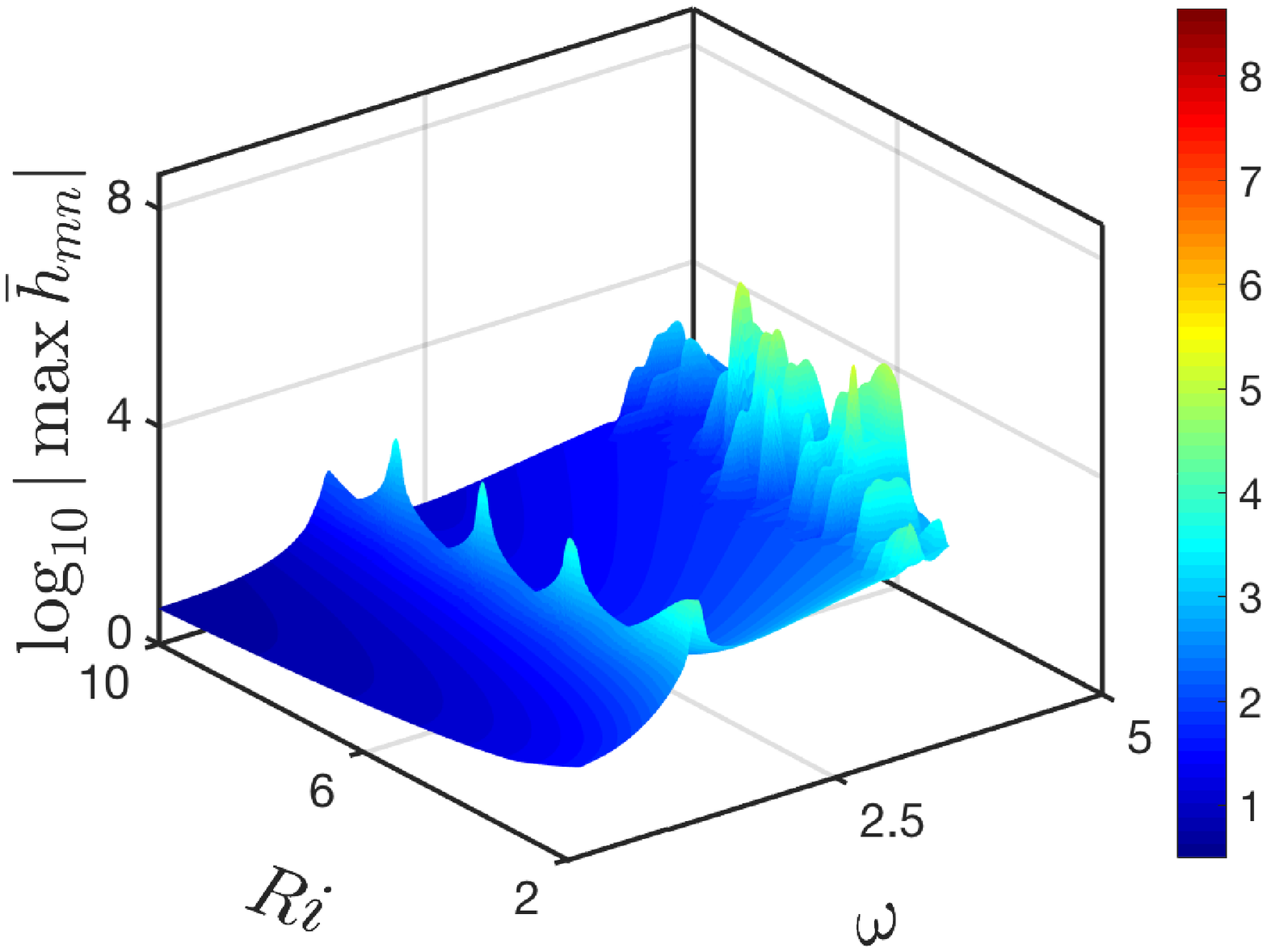}}
\subfigure[]{
\includegraphics[width=6.5cm,height=4.0cm]{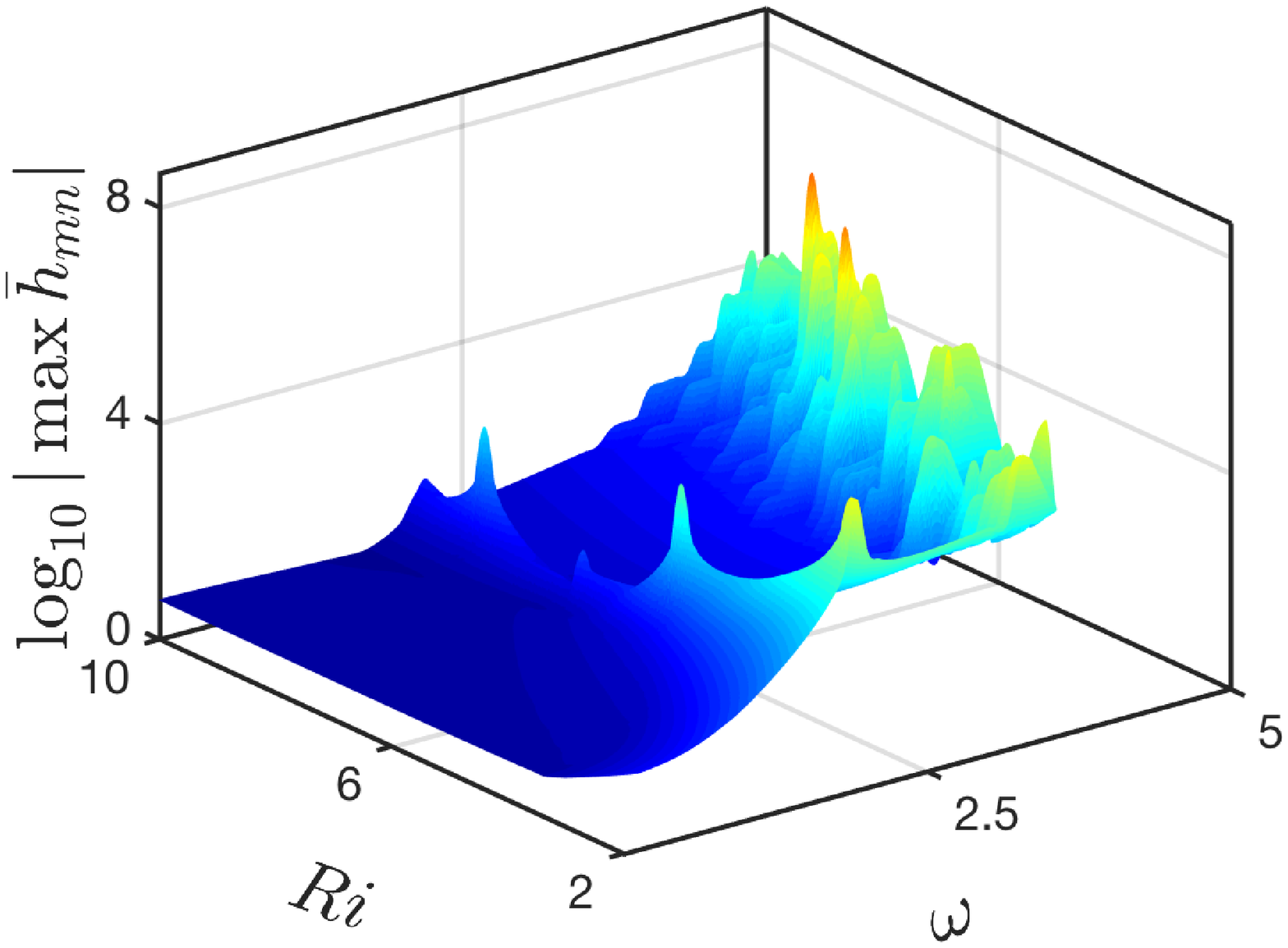}}\\
\subfigure[]{
\includegraphics[width=6.5cm,height=4.0cm]{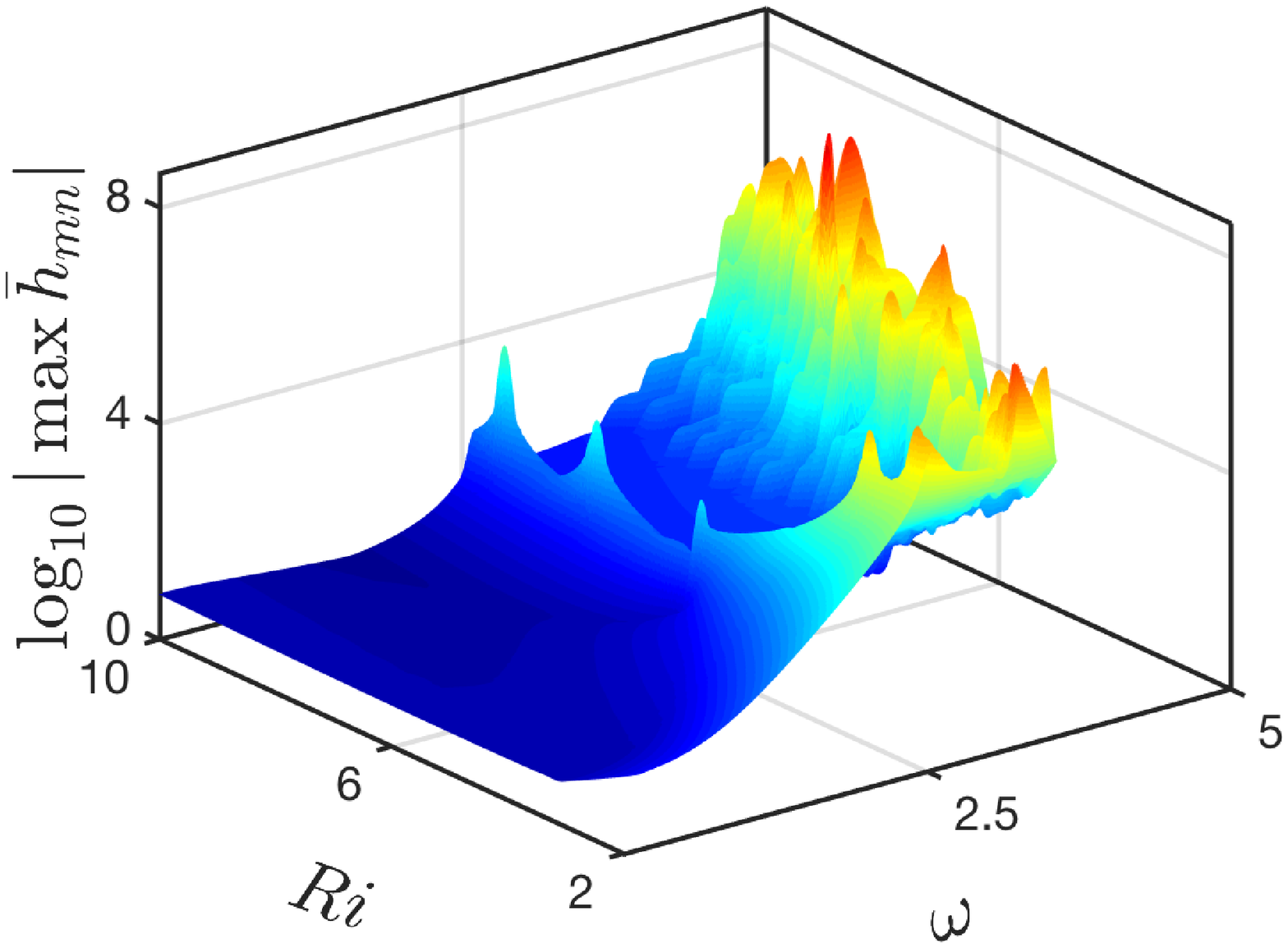}}
\subfigure[]{
\includegraphics[width=6.5cm,height=4.0cm]{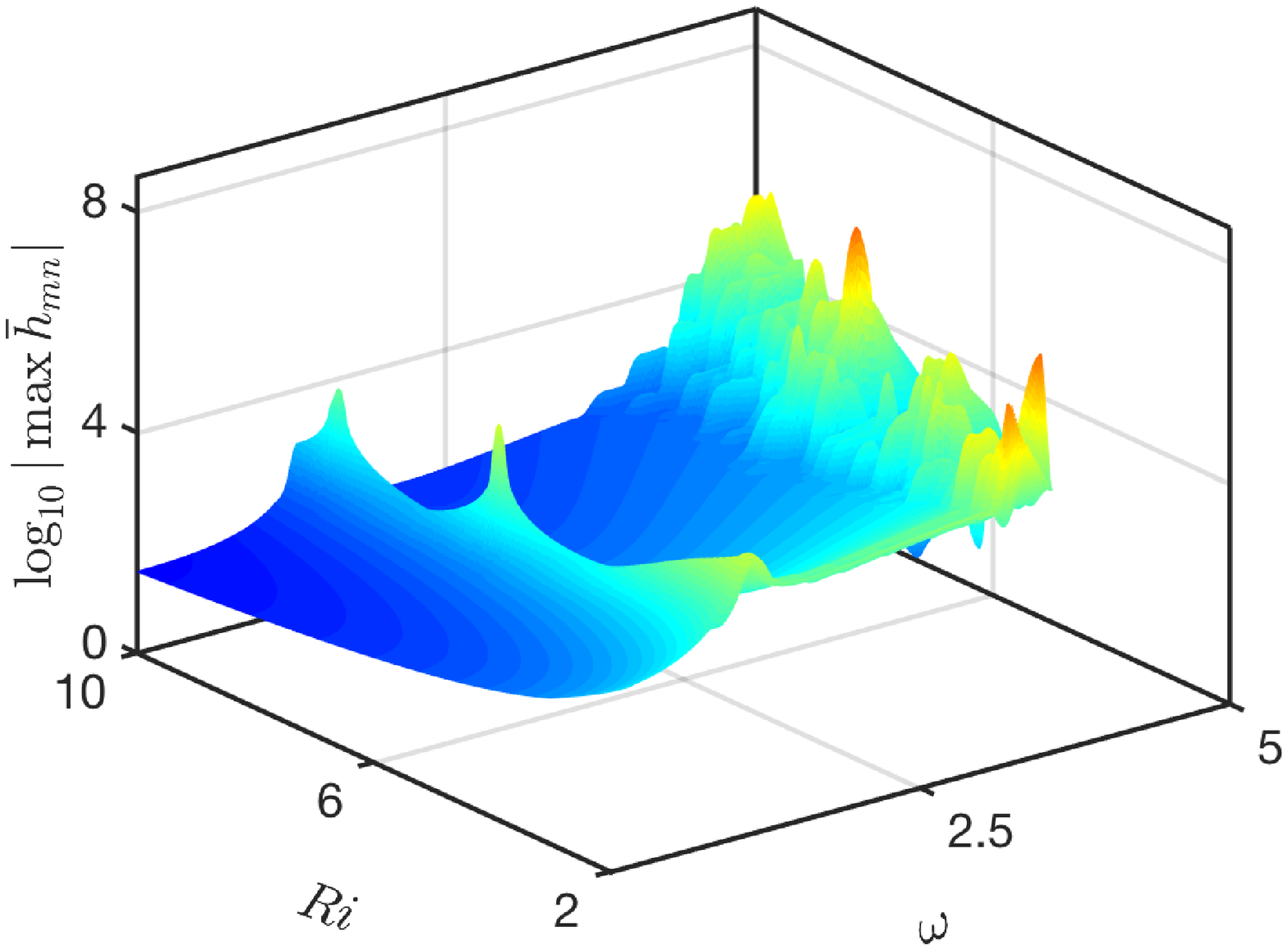}}\\
\subfigure[]{
\includegraphics[width=6.5cm,height=4.0cm]{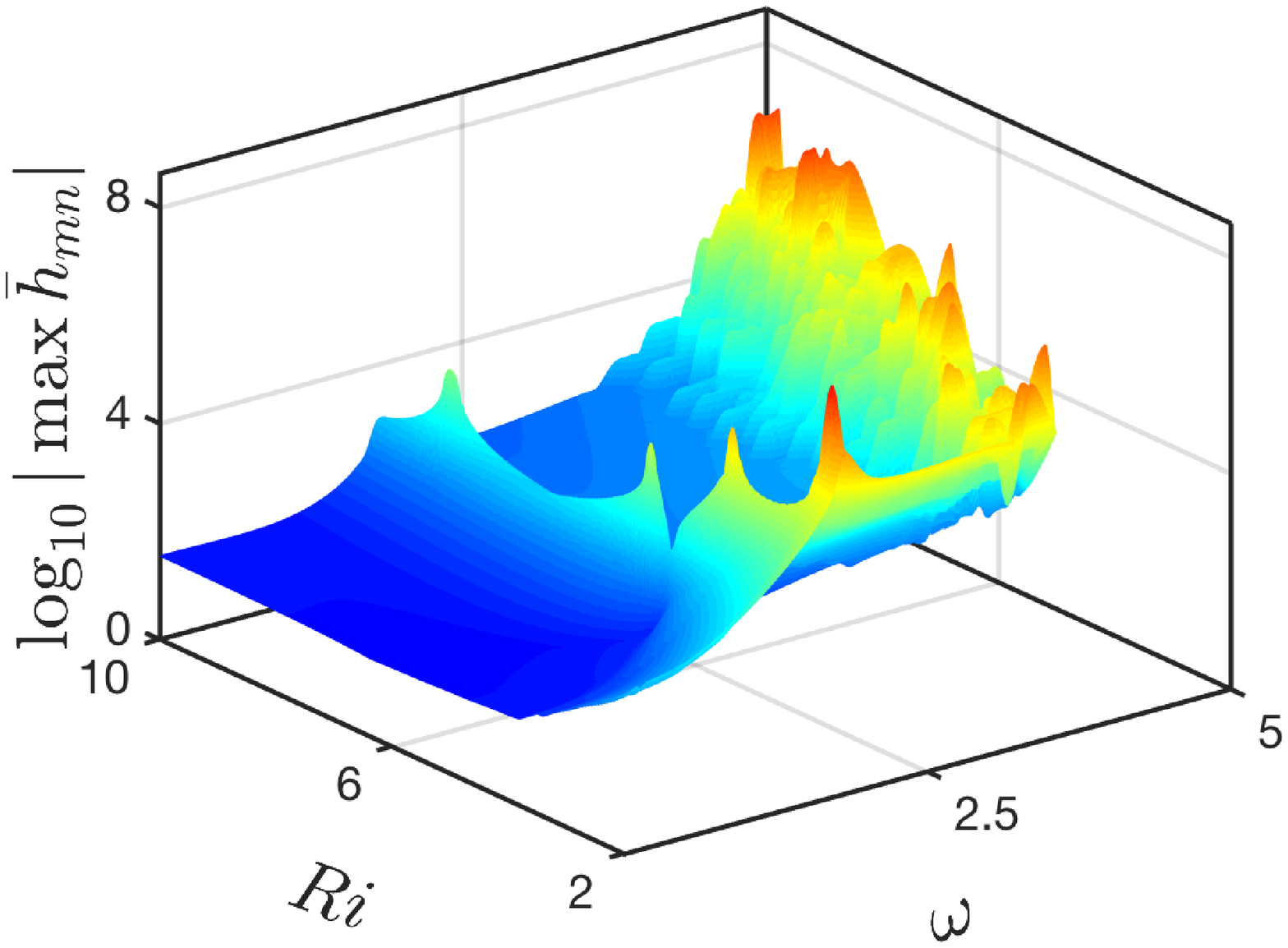}}
\subfigure[]{
\includegraphics[width=6.5cm,height=4.0cm]{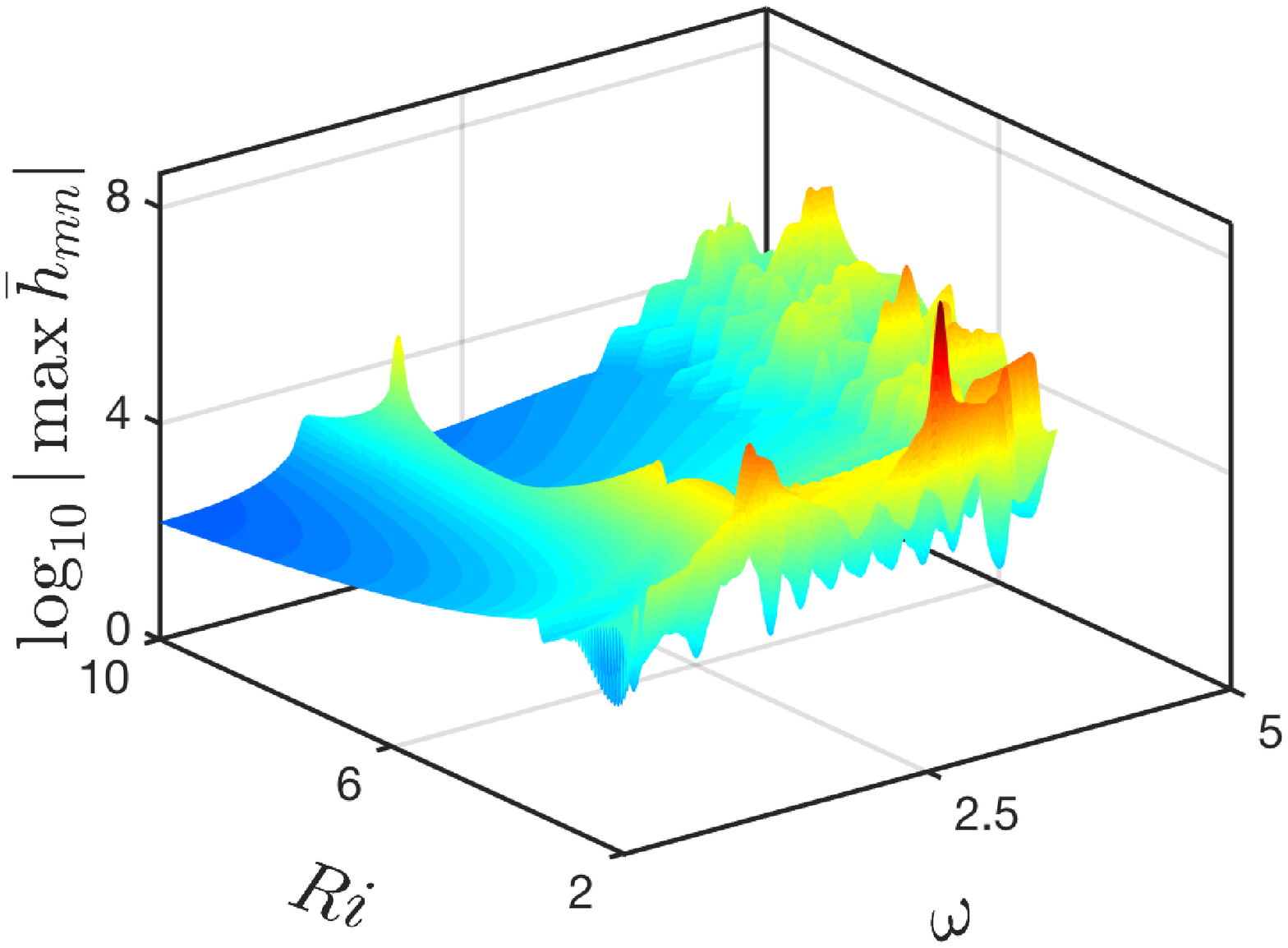}}\\
\caption{\small{The variation of $\log_{10}|\max\bar{h}_{mn}|$ in ($\omega$, ${\rm Ri}$)-plane for six mode pairs $(m,n)$: (a) (1,2), (b) (1,3), (c) (1,4),  (d) (2,3),  (e) (2,4), (f) (3,4). }
}
\label{3d_plots}
\end{figure}

Up till now, we analysed the evidence of  RTI for selected values of $(\omega, {\rm Ri})$ by exactly finding the resonating modes by the mode search method and also by graphically analyzing the resonance conditions~\eqref{eqn:triad_condition}. However, for wider range of parameters $(\omega, {\rm Ri})$ the mode search method is rather tedious. To avoid this adversity, one can directly use the fact that the  second-order solution $\bar{h}_{mn}(z)$ diverges at the resonance point on $(\omega, \rm Ri)$-plane due to the singularity of the associated linear operator $L_2^+$. In the following we look for the possibility of RTI by exploiting this fact. 

For each mode pair $(m,n)$ figure~\ref{3d_plots} shows the surface plot in $(\omega, {\rm
Ri},\log_{10}|\max\bar{h}_{mn}|)$-plane,  where $\omega\in[0.01, 5]$ and ${\rm Ri}\in[2,10]$. The colorbar is scaled between $0.5$ to $8.6449$ where the dark red denotes an upper bound of the colorbar and the blue a lower bound. Owing to the divergence of $\bar{h}_{mn}(z)$, $\log_{10}|\max\bar{h}_{mn}|$ attain large values as compared to the neighbouring points in the locations of $(\omega, {\rm Ri})$-plane where RTI might occur. Nevertheless, not all such $(\omega, {\rm Ri})$ values where $\log_{10}|\max\bar{h}_{mn}|$ achieve large values correspond to the exact RTI. To put it differently, large values or peaks in the surface plot may appear due to the near as well as the exact RTI. For all $(\omega, {\rm Ri})$ where $\bar{h}_{mn}(z)$ diverges, we check the resonance conditions~\eqref{eqn:triad_condition} in order to confirm that whether there is any possibility of an exact resonant triad formation.  

It is seen from figure~\ref{3d_plots}(a) that for $(m,n)=(1,2)$ the order of $\log_{10}|\max\bar{h}_{mn}|$ is less than or equal to $4$ for given range of $\omega$ and ${\rm Ri}$, which can also be seen from table~\ref{tab:resonance_table1}. It is verified that the peaks in the variation of $\log_{10}|\max\bar{h}_{mn}|$ occur due to the near RTI. It is worth recalling that for $(m,n)=(1,2)$, $\mathcal{O}(\Delta_k)$ and $\mathcal{O}(\Delta_\omega) \geq 10^{-4}$ for all $(\omega, {\rm Ri})$ which also implies that the exact RTI are not possible in this case, for the stratification profiles and the frequency of interacting primary modes considered in this paper.

By comparing all the panels (a--f) of figure~\ref{3d_plots} it is found that among all mode pairs 
$\max\limits_{1\leq m< n\leq4}{\log_{10}|\max\,\bar{h}_{mn}|}$ is attained when $(m,n)=(3,4)$ for $(\omega, {\rm Ri})=(2.61,2)$; and for this case the second-order solution $\bar{h}_{mn}(z)$ diverges and thereby leading to the exact resonant triad formation. Recall from table~\ref{tab:resonance_table1} that for  the above mentioned  parameters, $\Delta_\omega$ is indeed of order $10^{-8}$ for $(m,n)=(3,4)$ and hence giving rise to the exact RTI. 

\begin{table}
\centering
\begin{tabular}{|l|l|}
\hline
$(m,n)$ &$\omega_{\min}$ \\
\hline
(1, 3)&1.91\\
(1, 4)&2.41\\
(2, 3)&1.21\\
(2, 4)&1.71\\
(3, 4)&0.01\\
$(m, m)$ &0.01\\
 for $m$=1, 2, 3, 4 & \\
\hline
\end{tabular}
\caption{\small{A minimum frequency below which RTI are not possible.}}
\label{tab:omega_min}
\end{table}

For mode pairs $(1,4)$ and ($2,4)$, see panels~(c) and (e), there exist several ($\omega, \rm Ri$) pairs for which $\log_{10}|\max\,\bar{h}_{mn}|$ attains large values implying that the possibility of RTI at those points, see red peaks in panel (c, e). On the other hand, in the case of mode pair $(1,3)$, $|\max\,\bar{h}_{mn}|$ is very large ($\mathcal{O}(10^6)$) at two values of ($\omega, {\rm Ri}$) especially for (4.51, 8.2) and (4.21, 7.0), as compared to their neighboring points, see panel~(b). For these parameters $\Delta_\omega\approx 0$ and $\Delta_k\approx 0$ thereby giving rise to once again the exact RTI. 

Likewise, in the case of mode pair $(2,3)$ we have obtained some points in ($\omega, {\rm Ri}$) plane by observing the behaviour of $\log_{10}|\max\,\bar{h}_{mn}|$ where near RTI occur, see panel (d). At those points the existence of near RTI can be verified by the mode search method.

The above results show that for each considered mode pair, one can always define a minimum frequency $\omega= \omega_{\min}$ below which RTI are not possible for any value of Richardson number (${\rm Ri}$) considered in the present analysis, see table~\ref{tab:omega_min}. In other words, the operator $L_2^+$ is always non-singular for $\omega<\omega_{\min}$ that rules out the possibility of RTI.

\subsection{Self interaction}

\begin{table}
\centering
\scalebox{0.75}{
\begin{tabular}
{
>{$}c<{$} >{$}c<{$} >{$}c<{$} >{$}c<{$} >{$}c<{$} >{$}c<{$} >{$}c<{$} >{$}c<{$}}
{\rm Ri} & \omega & k_m  &\omega_r & k_r^{2\omega}&\Delta_k&\Delta_\omega& \max|\bar{h}_{mn}|
\\
\hline
\\
\multicolumn{8}{c}{$(m, n)=(1, 1)$}\\ \hline
\rowcolor[gray]{0.8}
9.0&0.01& 4.8617\times10^{-3} &0.02  &0.00972345 &5.31215\times10^{-8}&  2.1087\times10^{-7}& 2.3922e\times10^{4} \\
9.8&0.01 & 4.6862\times10^{-3} &0.02 &0.0093725&9.52855\times10^{-8}& 2.1570\times10^{-7}&  2.1425\times10^{4} \\
10.0&0.01 & 4.6452\times10^{-3}&0.02 &0.0092905&1.00899\times10^{-7}& 2.1892\times10^{-7}&  2.0676\times10^{4}  \\
 9.9& 0.01& 4.6656\times10^{-3}  &0.02 &0.00933123&2.9569\times10^{-8}& 2.2026\times10^{-7}& 2.0764\times10^{4}  \\
9.7& 0.01& 4.7071\times10^{-3}  &0.02 &0.00941431 &1.09867\times10^{-7}&  2.2363\times10^{-7}& 2.0885\times10^{4}  \\ \hline
\multicolumn{8}{c}{$(m, n)=(2, 2)$}\\ \hline
\rowcolor[gray]{0.8}
3.3&0.01 & 9.4669\times10^{-3}&0.02  &0.0189339&7.5857\times10^{-8}& 6.7452\times10^{-8}&  5.8746\times10^{5}\\
3.0&0.01 & 9.5576\times10^{-3}&0.02  &0.0191153 &8.6917\times10^{-8}&  6.7712\times10^{-8}&  6.6777\times10^{5}  \\
   3.2&0.01& 9.4974\times10^{-3} &0.02 &0.0189949&6.49842\times10^{-8} &  7.0318\times10^{-8}&  5.8787\times10^{5} \\
   3.1&0.01 & 9.5276\times10^{-3} &0.02 &0.0190554& 1.64842\times10^{-7}&  7.0365\times10^{-8}&  6.1380\times10^{5}  \\
   3.7&0.01& 9.3432\times10^{-3} &0.02 &0.0186866 & 1.53188\times10^{-7}&  7.4906\times10^{-8}&  4.5275\times10^{5}  \\
\hline
\multicolumn{8}{c}{$(m, n)=(3, 3)$}\\ \hline
\rowcolor[gray]{0.8}
3.0&0.01 & 9.9322\times10^{-3} &0.02&0.0198644 & 9.285\times10^{-9}&  1.0225\times10^{-8}&  1.1536\times10^{7} \\
3.4&0.01&9.9017\times10^{-3}&0.02 &0.0198034& 3.525\times10^{-8}&1.1058\times10^{-8}&  8.5357\times10^{6}   \\
3.1&0.01&9.9250\times10^{-3} &0.02&0.0198501 &9.223\times10^{-8}&  1.1736\times10^{-8}&  9.4670\times10^{6} \\
3.2&0.01&9.9176\times10^{-3}&0.02 &0.0198351&5.832\times10^{-8}&1.1984\times10^{-8}&  8.7605\times10^{6} \\
3.6&0.01&9.8846\times10^{-3}&0.02&0.0197692&2.310\times10^{-8}&  1.3052\times10^{-8}&  6.5528\times10^{6} \\
 \hline
\multicolumn{8}{c}{$(m, n)=(4, 4)$}\\ \hline 
\rowcolor[gray]{0.8}
4.7&0.01&9.9484\times10^{-3}&0.02 &0.0198968&2.868\times10^{-8}&  4.3298\times10^{-9}&  2.7053\times10^{7} \\
4.5&0.01&9.9550\times10^{-3}&0.02&0.0199101 &9.768\times10^{-8}&   4.4900\times10^{-9}&  2.8398\times10^{7}\\
4.6&0.01&9.9518\times10^{-3}&0.02 &0.0199035&5.076\times10^{-8}&  5.0025\times10^{-9}&  2.4414\times10^{7} \\
4.3&0.01&9.9612\times10^{-3}&0.02 &0.0199225&9.545\times10^{-8}& 5.4863\times10^{-9}&  2.5417\times10^{7}  \\
4.8&0.01&9.9449\times10^{-3}&0.02 &0.0198898 &3.319\times10^{-8}&  5.5337\times10^{-9}&  2.0329\times10^{7} \\
\hline
\end{tabular}}
\caption{\small{ 
Same as table~\ref{tab:resonance_table1} but for the self-interaction cases, i.e. interaction between $(m,m)$ mode pair.}
}
\label{self interaction table}
\end{table}

For self-interaction, i.e.~$m=n$ or $k_m=k_n$, the superharmonic system~\eqref{eqn:hmn} reduces to, 
\begin{align}
L_2^{+} h_{mm}(z) = A_{mm}(z),
\label{self interaction eqn}
\end{align}
where
\begin{align*}
L_2^+ \equiv  
 - &\left(-2\omega+  2\,k_m \bar{U}(z)\right)^2
(D^2- 4\, k_m^2)- 4 \,{\rm Ri}\, k_m^2\, ,
\label{eqn:Self_L+}
\\
A_{mm}(z) = 2\,k_m^2&\left[-\left(\bar{U}(z)-k_m/\omega\right)\left(\psi_m(D^2-k_m^2)D\psi_m-D\psi_m(D^2-k_m^2)D\psi_m\right)\right.\nonumber\\
&\left.
+{\rm Ri}_{0}\left(\rho_mD\psi_m-\psi_mD\rho_m\right)\right]. 
\end{align*}
In the case of RTI arising through the interactions of a primary mode with itself, singularity of  $L_2^+$ operator leads to the diverging spatial amplitude of the secondary mode, $h_{mm}(z)$. In the following we discuss RTI through the self-interactions of modes for different $\rm Ri$. We follow the same analysis as for $m\neq n$ to identify RTI for the self-interaction case $m=n$.

Table~\ref{self interaction table} is the same as table~\ref{tab:resonance_table1} but for the self-interaction case. It is evident from table~\ref{self interaction table} that both $\Delta_k$ and $\Delta_\omega$ are of order $\leq 10^{-7}$  leading to the exact RTI at $\omega=0.01$ and at various ${\rm Ri}$ listed in the table. In contrast to $m \neq n$ case,  RTI appear at smaller frequency  for each mode number ($1\leq m \leq 4$) and all Richardson number in the  range $[2,10]$.

\begin{figure}
\centering
\includegraphics[scale=0.26]{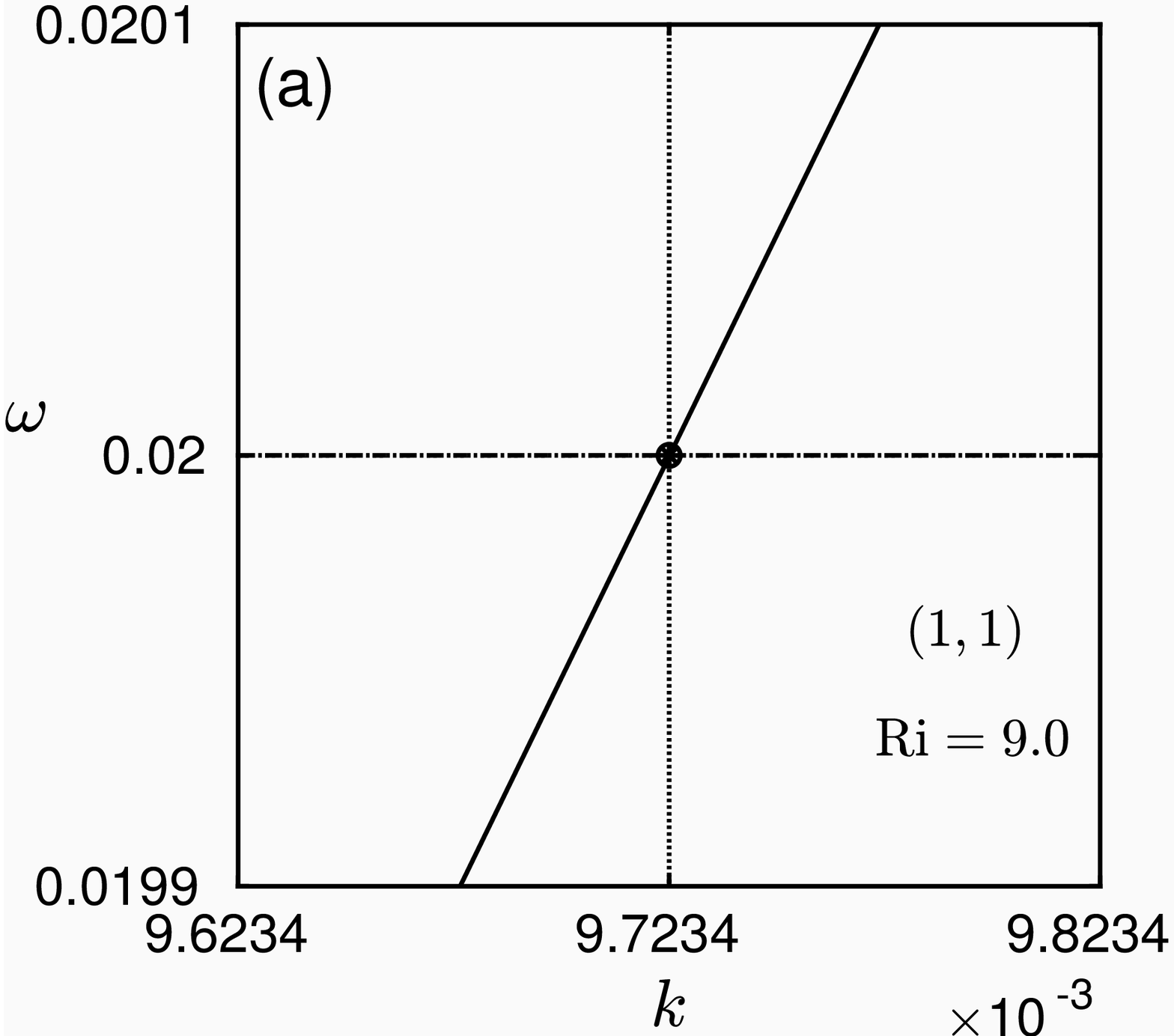}\qquad\qquad\quad
\includegraphics[scale=0.26]{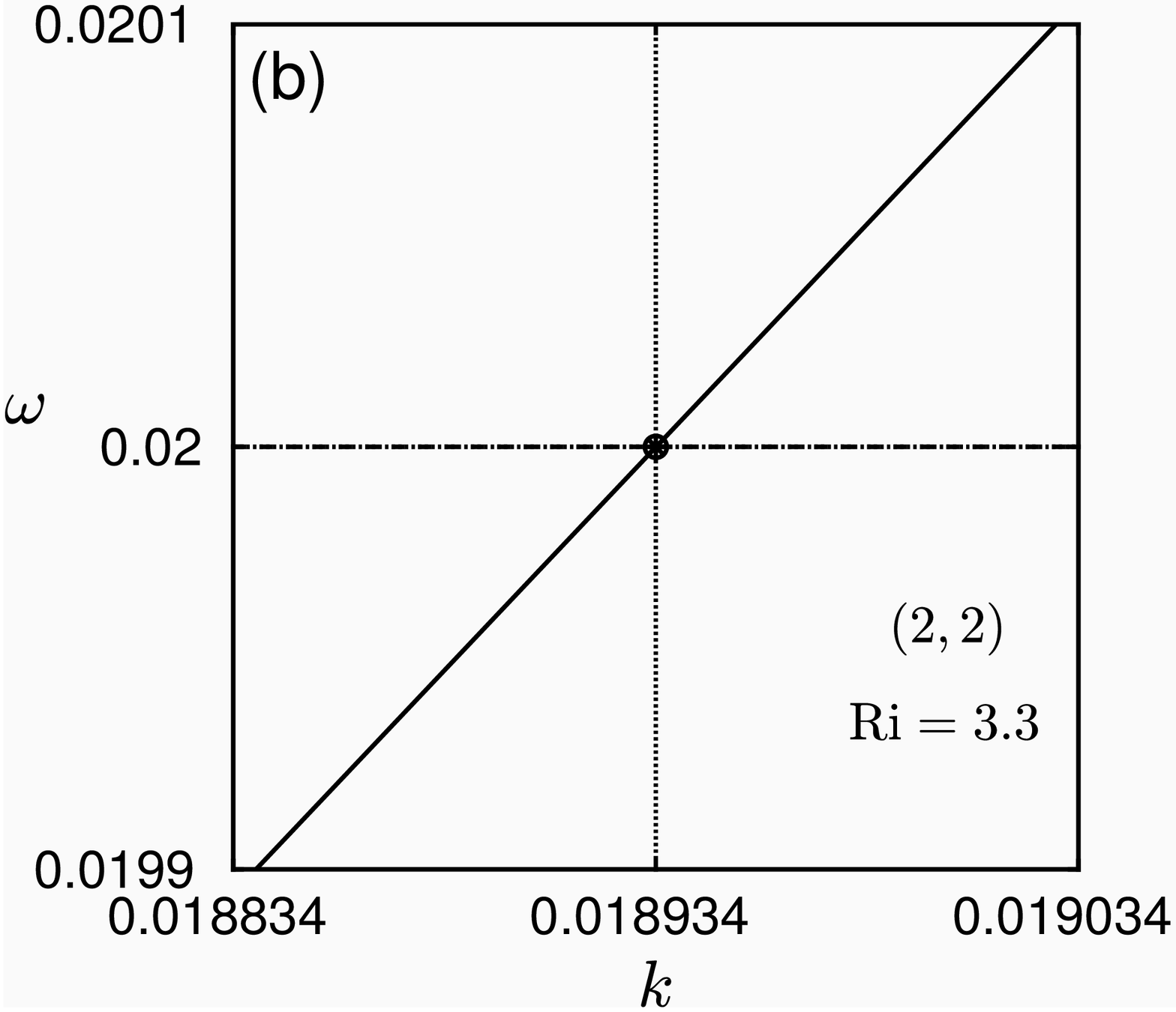}\\[1em]
\includegraphics[scale=0.28]{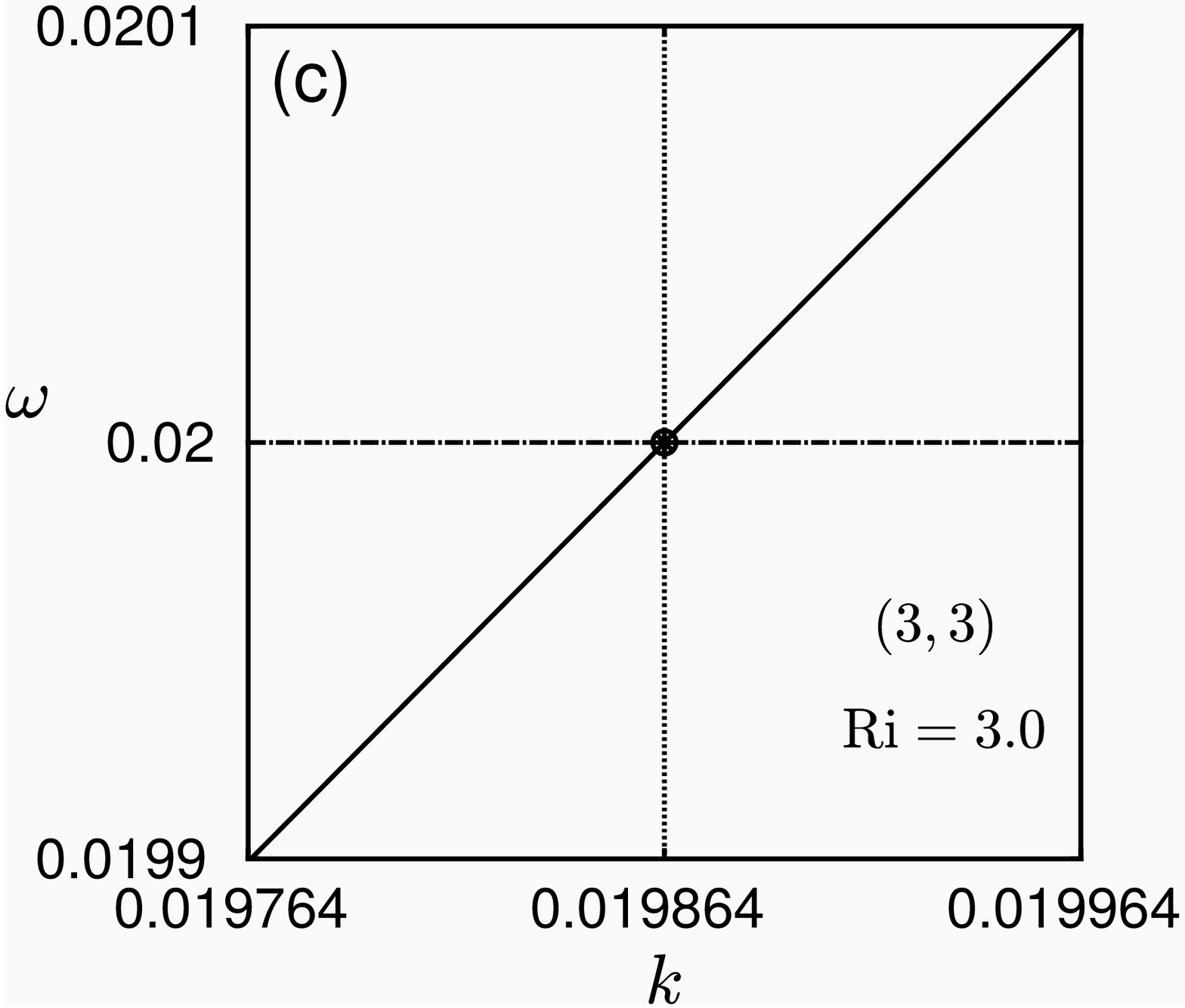}\qquad\qquad
\includegraphics[scale=0.28]{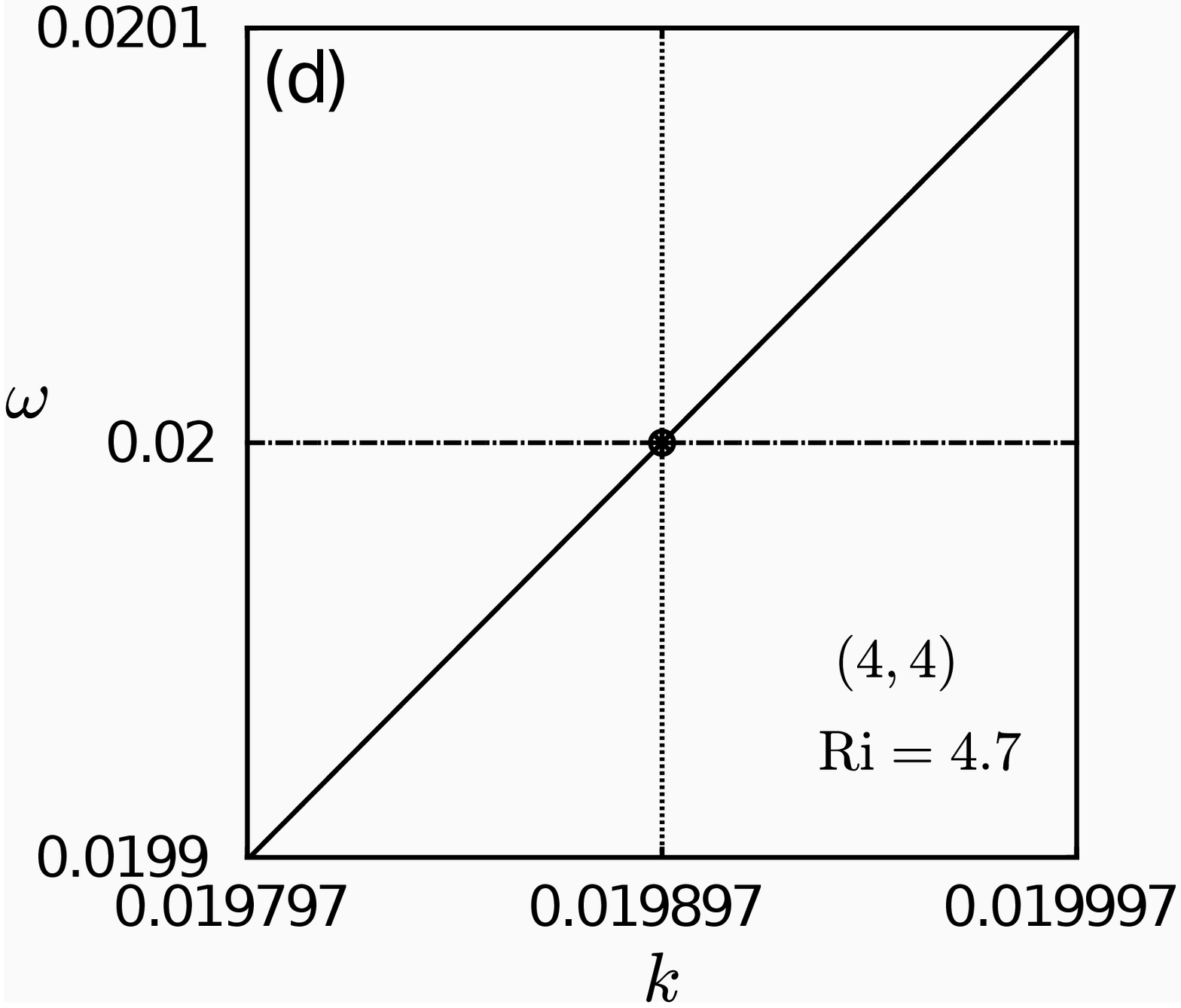}\\
\caption{\small{
Same as figure~\ref{2d plots} but for self-interaction case of RTI.} }
\label{self interaction 2d plots}
\end{figure}

For parameter values corresponding to the shaded row, we show the dispersion relation in figure~\ref{self interaction 2d plots}. Similar to figure~\ref{2d plots}, circle and star depict the points ($2k_m,\, 2\omega$) and ($2k_m,\,\omega_r$) which coincide ($\Delta_\omega=0$) with each other and hence verifying the presence of the exact RTI.

\subsection{Divergence surface for \texorpdfstring{$m=n$}{m eq n}}

\begin{figure}
\centering
\subfigure[]{
\includegraphics[width=6.5cm,height=4.0cm]{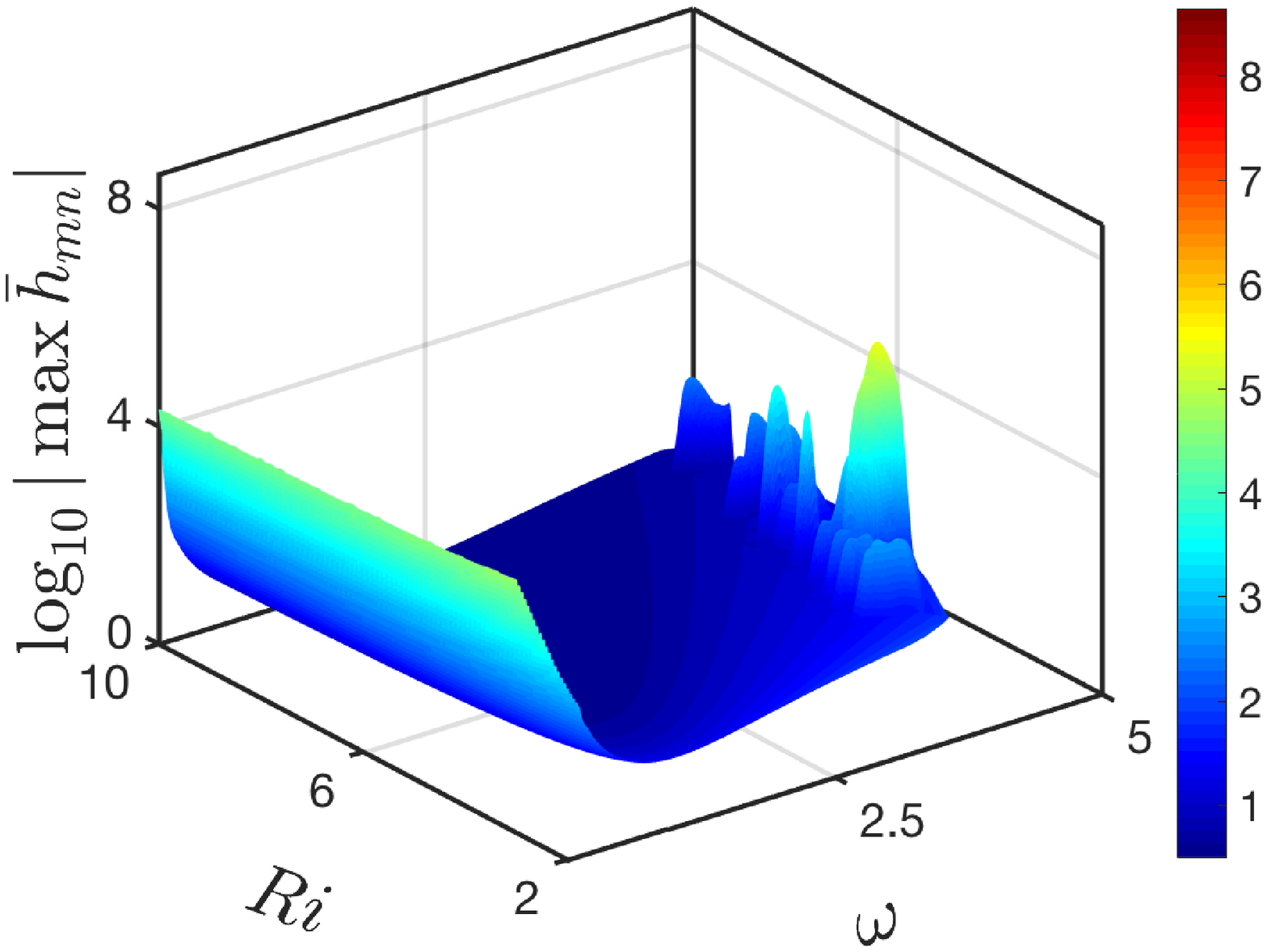}}
\subfigure[]{
\includegraphics[width=6.5cm,height=4.0cm]{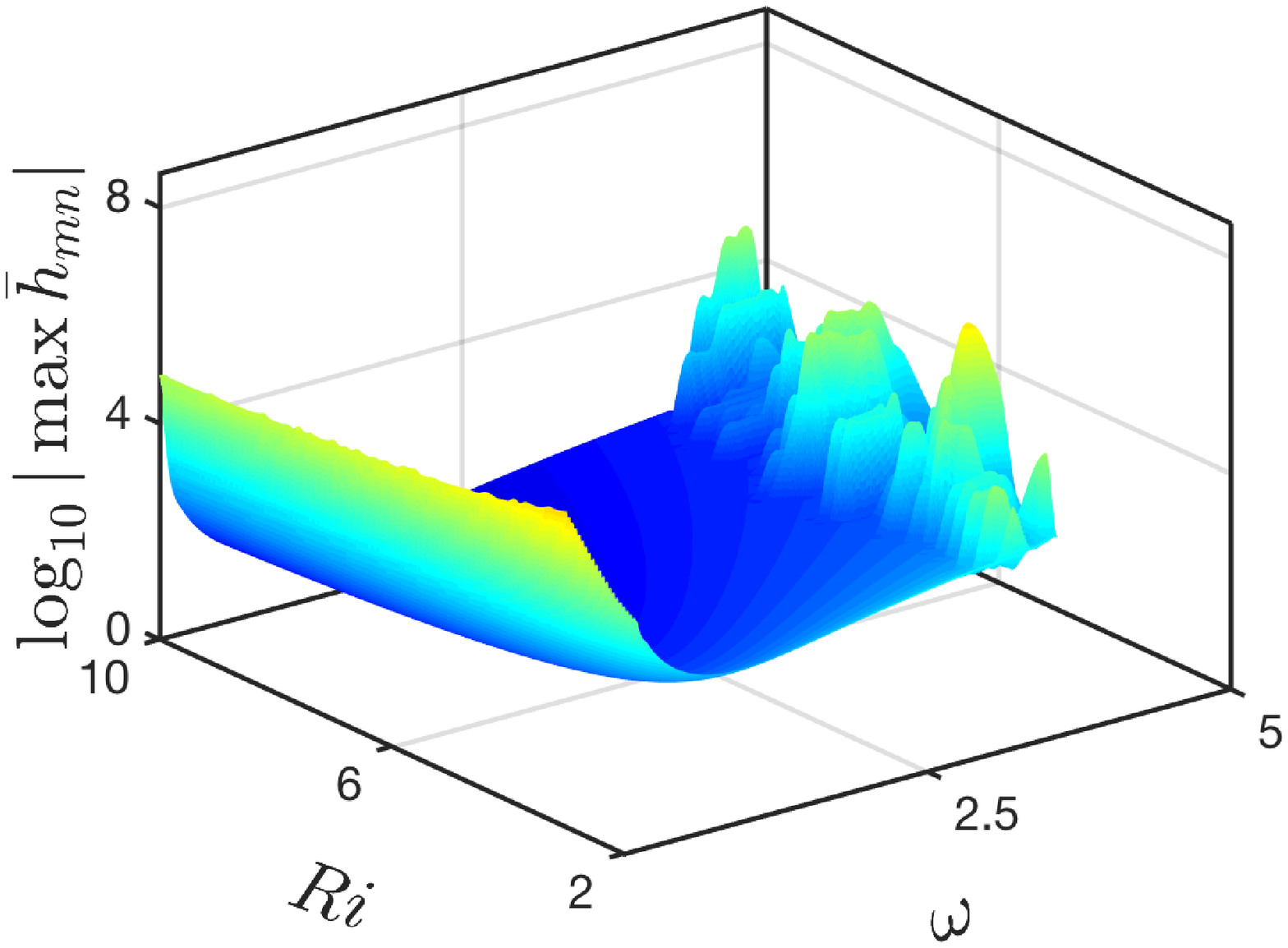}}\\
\subfigure[]{
\includegraphics[width=6.5cm,height=4.0cm]{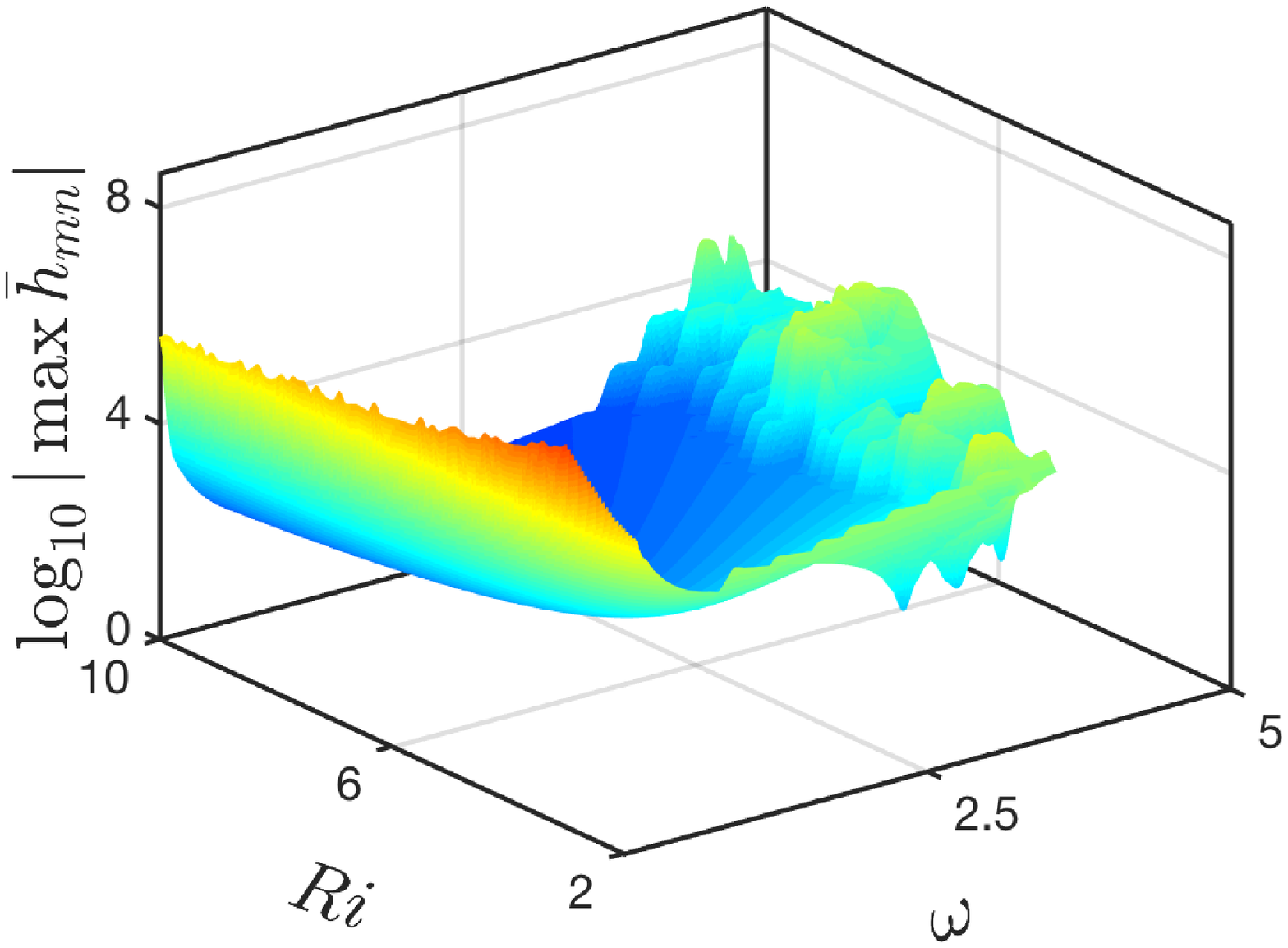}}
\subfigure[]{
\includegraphics[width=6.5cm,height=4.0cm]{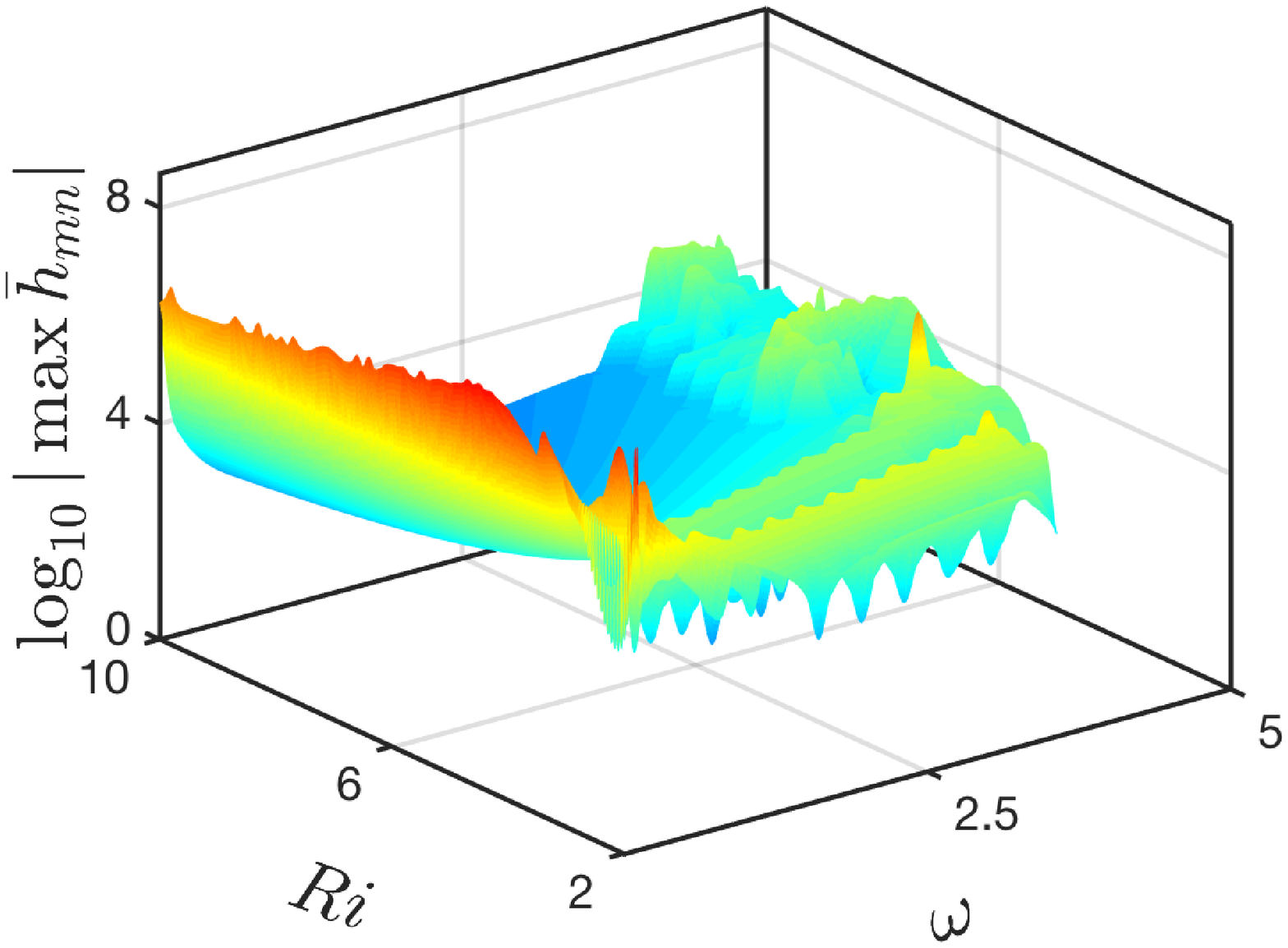}}\\
\caption{\small{
Same as figure~\ref{3d_plots} but for 
self-interacting modes: (a) $m=1$, (b) $m=2$, (c) $m=3$ and (d) $m=4$.
}}
\label{self_3d_plots}
\end{figure}

In order to probe RTI for wider range of parameters we show the three dimensional surface plots, similar to figure~\ref{3d_plots}, and look for the possibility of resonance when a wave mode with wavenumber $k_m$ with frequency $\omega$ interacts with itself. Panels (a) to (d) of figure~\ref{self_3d_plots} correspond to $m=1$, $2$, $3$ and $4$, respectively. The quantities $\Delta_\omega$ and $\Delta_k$ are computed for all those points where $\bar{h}_{mn}(z)$ diverges, or $\log_{10}|\max\,\bar{h}_{mn}|$ has a peak, in order to check how accurately the resonant conditions are satisfied for a given ($\omega, {\rm Ri}$). 

By analyzing a resonance triad formed by  self-interactions of modes, it is shown that  RTI exist for different uniform stratifications (i.e. for all ${\rm Ri}$) considered here, at frequency $\omega=0.01$. The order of the parameter $\Delta_\omega$ (or $\Delta_k$) classifies the triads as near or exact resonant triads.

In the case of exact RTI, the strength of a resonant triad depends on  the accuracy of the resonance conditions, i.e. strength of resonance triad is inversely proportional to $\mathcal{O}(\Delta_\omega)$ or $\mathcal{O}(\Delta_k)$. It is interesting to note that the strength of exact resonance triad  increases with the mode number of the self-interacting mode, i.e. the strongest resonance occurs for $m=4$. This can also be seen from the values of $\Delta_\omega$ displayed in $7^\mathrm{th}$ column of table~\ref{self interaction table}. By comparing panels~(a) to (d) of figure~\ref{self_3d_plots}, it is evident that  the surface plots exhibit more peaks as mode number increases and hence give rise to more possible points of resonances in ($\omega,\, {\rm Ri}$)-plane. Thus, larger mode number implies  more possible positions of exact resonances, for example, the number of possible resonant points in $(\omega, {\rm Ri})$-plane is minimum for $m=1$ and maximum for $m=4$. At those possible resonance points the type of RTI (near or exact) is determined by the mode search method.

\section{Conclusion}
\label{sec:Conclusion}

In this paper, we have discussed an interaction of internal gravity waves in a stably stratified uniform shear flow bounded between two infinite horizontal parallel plates. In particular, we have focused on RTI formed between the two primary modes of the same frequency $\omega$ and a superharmonic mode of frequency $2\omega$. The dispersion relation $\mathcal{D}(\omega,k; \rm Ri)=0$ has been solved analytically as well as numerically. Through the dispersion relation various wavenumbers $k_m=k_m(\omega)$ as well as various frequencies $\omega_m=\omega_m(k)$ are calculated, where $m$ is a positive integer. 

The second order solution corresponding to the interaction between two primary modes with mode numbers $m$ and $n$ constitutes two different waves---one is time dependent mode at frequency $2\omega$ with wavenumber $k_m+k_n$, referred to as superharmonic mode and another is time independent mode with wavenumber $k_m-k_n$, referred to as subharmonic mode. Spatial amplitude of the superharmonic mode $\bar{h}_{mn}(z)$ blows up when the resonance conditions are satisfied. Thus the existence of a resonance triad has been found to be associated with the divergence of the second-order solution. Here we consider RTI for different uniformly stable stratifications by varying the square of dimensionless buoyancy frequency ($N^2/N_0^2$) as well as the  local Richardson number ($\rm Ri$). In addition, for a given frequency and the Richardson number, the possibility of  RTI involving the primary modes with mode numbers $m$ and $n$ at frequency $\omega$ and a superharmonic mode at frequency $2\omega$ has also been explored in terms of the resonance triad conditions, i.e.~$k_1+k_2=k_3$ and $\omega_1+\omega_2=\omega_3$.

It has been established that for a given $(m,n)$ mode pair, the resonance conditions are indeed met at some specific peaks---referred to as diverging second-order solutions---of the surface plot in $(\omega, {\rm Ri},\log_{10}|\max\,\bar{h}_{mn}|)$-plane. The diverging peaks are those peaks where the spatial amplitude $\bar{h}_{mn}(z)$ of the superharmonic mode is divergent implying the existence of a resonance triad. In the self-interaction case (i.e. in the case of $m=n$), the divergent peaks occur at very small frequencies in comparison with the interaction between two different modes $(m\neq n)$. The intensity of the divergent peaks increases with the mode number in both the cases ($m\neq n$ and $m=n$); consequently the range of parameters $(\omega,\, {\rm Ri})$ increases where the amplitude of superharmonic waves has a large magnitude. It has been shown that the resonant triad exists for almost all values of ${\rm Ri}\in [2,10]$ and at frequency as small as $\omega=0.01$ for self-interacting modes $(m=n)$.  

The superharmonic solution diverges when any of the  following three conditions are satisfied: (i) $k_m + k_n= k_r^{2\omega}$, where $k_r^{2\omega}$ is the wavenumber of a mode $r$ having frequency $2\omega$, (ii) $2\omega= \omega_r$, where $\omega_r$ is the frequency of a mode $r$ at wavenumber $k_m+k_n$ and (iii) the right-hand side $\bar{A}_{mn}(z)$ of~\eqref{eqn:hmn} must be non-orthogonal to the adjoint eigenfunction associated with the homogeneous problem. While the first two conditions are the standard triadic resonance conditions, the last is the Fredholm alternative. Markedly, the last condition is always fulfilled when either of the first two is fulfilled (with a reasonably good accuracy). 

The criteria for exact RTI involving the modes $m$ and $n$ having frequency $\omega$ are $\Delta_k=|k_m + k_n - k_r^{2\omega}|=0$ and $\Delta_\omega=|2\,\omega - \omega_r|=0$. Notably, $\bar{h}_{mn}(z)$ diverges when $\Delta_k$ and $\Delta_\omega$ are exactly zero which is the case when the associated homogeneous equation has a non-trivial solution. Note that $\Delta_k$ and $\Delta_\omega$ are zero cannot be ascertained in the numerical solutions. Thanks to the above fact, due to which we could find the solution $\bar{h}_{mn}(z)$ and hence an integer $r$ in order to determine $k_r^{2\omega}$ as well as $\omega_r$. 
In contrast to the mode search method based only on the triad conditions, the superharmonic solution $\bar{h}_{mn}(z)$ allows us to determine the positions of resonance triads for the full range of $(\omega, {\rm Ri})$-plane more efficiently. Note that the present nonlinear problem of RTI has been tackled analytically and has also been verified numerically. The evidence of the resonant triad gives us an understanding about the energy transfer among modes of different wavenumbers. The present work can be extended in the future for more realistic scenarios where viscosity, earth rotation and diffusivity play an important role. Moreover, the present study would serve as a benchmark for several other stratified problem, for instance, multi-layer stratified flows, compressible stratified flows, etc.

\section*{Acknowledgment}

The authors thank to Dr.~Anubhab Roy for some discussions. 
P.S. thanks to Prof.~Manuel Torrilhon for giving many constructive suggestions in Mathematica. 
L.B. acknowledges the NBHM for financial support and
P.S. acknowledges the IIT Madras for a New Faculty Seed Grant (MAT/16--17/671/NFSC/PRIY).

\bibliography{refer_used}

\end{document}